\documentclass{aa}
\usepackage{graphicx}
\usepackage{txfonts}
\usepackage{amsmath}

\usepackage{kantlipsum} % Para salto de linea bien
\usepackage{soul}

\newcommand{\codename}{SICON}
\newcommand{\codedescription}{Stokes inversion based on convolutional neural networks}

\usepackage[colorlinks=true]{hyperref}
\hypersetup{citecolor= blue, linkcolor=blue, urlcolor=blue}
\usepackage{color}

\begin{document} 

   %\title{Three-dimensional spectropolarimetric inversions:\\a machine learning approach}
   %\title{Deep Learning Inversion of Stokes Profiles}
   \title{Stokes inversion based on convolutional neural networks}
   \author{A. Asensio Ramos
          \inst{1,2}
          \and
          C. J. D\'iaz Baso\inst{3}
          }

   \institute{Instituto de Astrof\'isica de Canarias, C/V\'{\i}a L\'actea s/n, E-38205 La Laguna, Tenerife, Spain
   \and
   Departamento de Astrof\'isica, Universidad de La Laguna, E-38206 La Laguna, Tenerife, Spain
   \and
   Institute for Solar Physics, Dept. of Astronomy, Stockholm University, AlbaNova University Centre, SE-10691 Stockholm Sweden
              % \email{cdiazbas@iac.es}
             }

   \date{}
   
   \titlerunning{}

% \abstract{}{}{}{}{} 
% 5 {} token are mandatory
 
  \abstract
   {Spectropolarimetric inversions are routinely used in the field of solar physics for 
   the extraction of physical information from observations. The application to 
   two-dimensional fields of view often requires the use of supercomputers with parallelized 
   inversion codes. Even in this case, the computing time spent on the process is still very large.}
   {Our aim is to develop a new inversion code based on the application of convolutional
   neural networks that can quickly provide a three-dimensional cube of thermodynamical and magnetic 
   properties from the interpreation of two-dimensional maps of Stokes profiles.}
   {We trained two different architectures of fully convolutional neural networks.
   To this end, we used the synthetic Stokes profiles obtained from two snapshots of three-dimensional 
   magneto-hydrodynamic numerical simulations of different structures of the solar atmosphere.}
   {We provide an extensive analysis of the new inversion technique, showing that it infers the thermodynamical 
   and magnetic properties with a precision comparable to that of standard inversion techniques. However,
   it provides several key improvements: our method is around one million times faster, it returns
   a three-dimensional view of the physical properties of the region of interest in geometrical height, it provides 
   quantities that cannot be obtained otherwise (pressure and Wilson depression) and the 
   inferred properties are decontaminated from the blurring effect of instrumental point 
   spread functions for free. The code, models, and data are all open source and available for
   free, to allow both evaluation and training.}
   {}


   \maketitle
%

%________________________________________________________________

\section{Introduction}
%The empirical determination of magnetic fields and thermodynamic properties of 
%different structures in the solar atmosphere is still crucial for our understanding of many 
%complex processes inside and outside magnetic active regions. 
Although numerical 
magneto-hydrodynamic (MHD) simulations of the solar atmosphere (especially
in the photosphere) are already in an advanced state 
\citep[see, e.g.,][]{Rempel12,Cheung10,Cheung2018_flare}, 
we continue to rely on the interpretation 
of spectropolarimetric signals with relatively simple empirical parametric 
models to determine magnetic fields and thermodynamic properties of
different structures in the solar atmosphere. The parametric 
nature of these empirical models makes it straightforward to tune the model parameters until
a good fit is obtained to the signals, so that it is relatively simple to extract the relevant information.

The inference of atmospheric parameters from Stokes profiles is 
done by proposing a forward model $\mathbf{I}(\boldsymbol{\theta},\lambda)$ that describes
the spectral shape of the four synthetic Stokes profiles depending on a set of model 
parameters $\boldsymbol{\theta}$. In general, fitting the model for the Stokes parameters
to the observations is a nonlinear nonconvex inverse problem that is plagued with
problems: difficulty in convergence, ambiguities as a consequence of the potential presence 
of several local minima (also extended minima in the hyperspace of parameters)
and computational problems due to the complexity of the
forward problem. In recent decades several models of increasing
complexity have been
proposed. The simplest ones are based on the Milne-Eddington atmosphere, which provides an 
analytical solution to the radiative transfer equation and results in an simple functional form
for the emerging Stokes parameters \citep{Auer1977,landi_landolfi04}. More complicated 
models require the presence of 
gradients along the line-of-sight (LOS). They were first implemented by \cite{sir92} in
the code ``Stokes Inversion based on Response functions'' (SIR) and it was
based on the assumption of local thermodynamical equilibrium (LTE) plus that of
hydrostatic equilibrium in the atmosphere. SIR describes the depth stratification of
the atmosphere in terms of a few nodes (their position and number are hyperparameters
that are typically selected by 
the user) and the rest of the stratification is obtained by a smooth interpolation. This
type of approach has been reimplemented later in many other codes, some of them
fundamentally similar like SPINOR \citep{Frutiger2000}, Helix+ \citep{Lagg2004} or VFISV 
\citep[Very Fast Inversion of the Stokes Vector;][]{Borrero2011}.
Other codes have relaxed some of the initial hypothesis of SIR. Examples include
NICOLE \citep{socas_navarro_2015} which  
relaxes the hypothesis of LTE and can extract information from strong chromospheric lines, 
STiC \citep{Jaime2018_Stic} which deals with lines with strong partial redistribution, needed
for very strong lines forming in the upper chromosphere or transition region, or Hazel \citep{Asensio2008}
which deals with lines whose polarization is controlled by scattering polarization, the
Hanle and Zeeman effects but in simplified model atmospheres.

In recent years, more than two decades after the publication of the SIR code, 
we are witnessing a bright new era in the development of inversion codes, with
new ideas all around. This is a consequence of several facts: i) the availability of 
more computing power, ii) the development and application of new techniques and algorithms and iii)
the imminent advent of the 4m class telescopes in solar physics with
DKIST \citep[Daniel K. Inouye Solar Telescope;][]{DKIST2012} and the European
Solar Telescope 
\citep[EST;][]{EST2013}
which will produce data of unprecedented quality and at an unprecedented rate.
From our perspective, one of the first examples of
this development was produced by \cite{vannoort12}, who proposed a spatially coupled inversion
code in which the point spread function (PSF) of the telescope couples together
many pixels in the field-of-view (FOV). The coupling produces that the 
inversion problem becomes global in the FOV and the model atmosphere for 
all pixels have to be simultaneously inferred.
% \textbf{To take into account  the point spread 
% function (PSF) of the telescope in the inversion problem, \cite{vannoort12} 
% proposed a spatially coupled inversion code where the model atmosphere for 
% all pixels have to be simultaneously inferred as they are coupled by the PSF.}
The solution of the optimization
problem via a Levenberg-Marquardt (LM) algorithm required the calculation of a Hessian
matrix that couples all pixels. Because the PSF is spatially compact, the Hessian
matrix has some sparsity structure and \cite{vannoort12} succeeded in exploting
this property to invert it, as required by the LM algorithm. However, 
the depth stratification of the physical quantities is
parameterized by a limited number of nodes to make the problem tractable.
With this approach, the results are decontaminated from 
the blurring effect of the PSF of the telescope, thus
avoiding the use of ad-hoc stray light contaminations, that often complicates
the interpretation of the results. A simpler and, in our personal view, also
more controllable approach was proposed
by \cite{ruizcobo_asensioramos13} \citep[see also][]{Quintero2015} in which 
deconvolution and inversion is separated in two processes. First 
the spectropolarimetric data is deconvolved
from the telescope PSF using a regularization based on the application of
principal component analysis (PCA) to the Stokes profiles. Then the inversion proceeds
pixel by pixel as in the usual case. This approach leads to a much simpler and 
computationally efficient workflow.

Another step forward was the application of the compressed sensing theory \citep{candes06} to
spatially regularize the maps of inverted physical
properties \citep{Asensio2015}. This approach takes advantage of the presence of spatial
correlation in the maps and effectively reduces the number
of unknowns that must be inferred. The code developed by \cite{Asensio2015}
uses a first-order proximal projection algorithm for the optimization of the sparsity-regularized
problem. On the downside, precisely the use of first-order optimization techniques 
produces slow convergence. 

Arguably the latest breakthrough in the field of inversion codes has been MASI \citep[MHD-Assisted 
Stokes Inversion;][]{Riethmuller2017}. This code uses a snapshot of an
MHD simulation as a lookup table when carrying out inversions. It works  
by comparing the observed Stokes parameters in every pixel of the FOV with
those of a properly degraded snapshot of an MHD simulation. The 3D cube of
physical properties associated with the observations is recovered by stacking
the columns of the simulation that better fits the observed Stokes profiles.
This 3D cube is obviously not in equilibrium so that it is used as an initial
condition in an MHD simulation code until a new equilibrium is found. The procedure
is then iterated until an eventual convergence. The resulting inversion code,
a result of the combination of a 3D MHD simulation module and a Stokes synthesis
module, is computationally demanding. \cite{Riethmuller2017} could only
iterate the process twice, though showing clear hints of convergence. In the
eventual convergence, one would find a 3D model of the region of interest that also fulfills the 
MHD equations.

Motivated by the recent enormous success of deep neural networks \citep[DNN;][]{Goodfellow2016}, we 
propose in this work to leverage DNN to carry out three-dimensional inversions 
of the solar atmosphere under the assumption of LTE for Hinode-like
observations. Our approach is termed \codename\ (standing for \codedescription).
We defer for the future the study of a neural approach
for the inversion of spectropolarimetric data at the diffraction limit of
a 4m class telescope like DKIST or EST and/or including non-LTE effects.
Data-driven approaches using DNNs have already been applied to solar
physics, demonstrating striking capabilities to solve problems that could not
be solved otherwise or with a much improved precision and/or speed. For instance, we 
managed to infer transverse velocities from pairs of consecutive 
images \citep[DeepVel;][]{Asensio2017}, which allowed us to identify 
small-scale vortices in the solar atmosphere that last for a few minutes and with
sizes on the order of a few hundred kilometers, something
impossible with methods based on local correlation tracking \citep{NovemberSimon_1988}. 
We have also applied them to
compensate for the blurring effect of the telescope and the Earth atmosphere
\citep{Enhance18,DeepMFBD18}. \cite{Illarionov18} applied DNNs for the automatic
segmentation of coronal holes with great success.
The advances in the field of deep learning suggests it is 
timely to investigate what are the prospects of neural networks in the field
of spectropolarimetric inversions. In this paper
we leverage convolutional neural 
networks \citep[CNN;][]{LeCun1998} that can easily exploit spatial information and 
the ability to train really deep neural networks that can approximate very nonlinear mappings.

Finally, we point out that the application of neural networks (multi-layer perceptrons; MLP) 
for the inversion of Stokes profiles is not
new. \cite{Carroll2001} already proposed their use for simple Milne-Eddington
inversions and concluded that they were able to obtain physical parameters
without any optimization once the neural networks were trained. As additional
advantages, they showed their speed, noise tolerance and stability. This
was later verified by other works \citep{socas_nn_03,socas_nn_05}. 
\cite{Carroll2008} later expanded their original work to use MLPs to
infer the depth stratification in a geometrical height scale of the temperature, velocity and
magnetic field vector, as their network was trained with a quiet Sun simulation \citep{vogler05}.
The application of the neural network pixel by pixel allowed them to recover
a tomographic view of the FOV by recombining all individual line-of-sight stratifications. 
More recently, \cite{osborne2019} has shown
how invertible neural networks \citep[INNs;][]{Ardizzone2018} can be successfully applied to capture 
degeneracies and ambiguities in the inference of thermodynamic parameters of
plane-parallel chromospheres of flaring regions.

Given the enormous flexibility in the architecture of DNNs, we follow in
this paper a double-blind strategy.
Both authors, starting from the same training dataset that is described in the following, studied
different architectures independently and ended up with two
proposals\footnote{Both architectures have been subject to simple ablation studies 
in which we tested the quality of the prediction by modifying and removing some 
features of the model.}. 
The general advantages and disadvantages
of both architectures (obviously many other architectures are possible) are
compared. Both neural networks, with their corresponding trained weights used
in this work, can be downloaded from our repository\footnote{\url{http://github.com/aasensio/sicon}}.
We anticipate that both architectures give very similar results.

\begin{figure*}[!ht]
    \centering
    \includegraphics[width=0.35\textwidth]{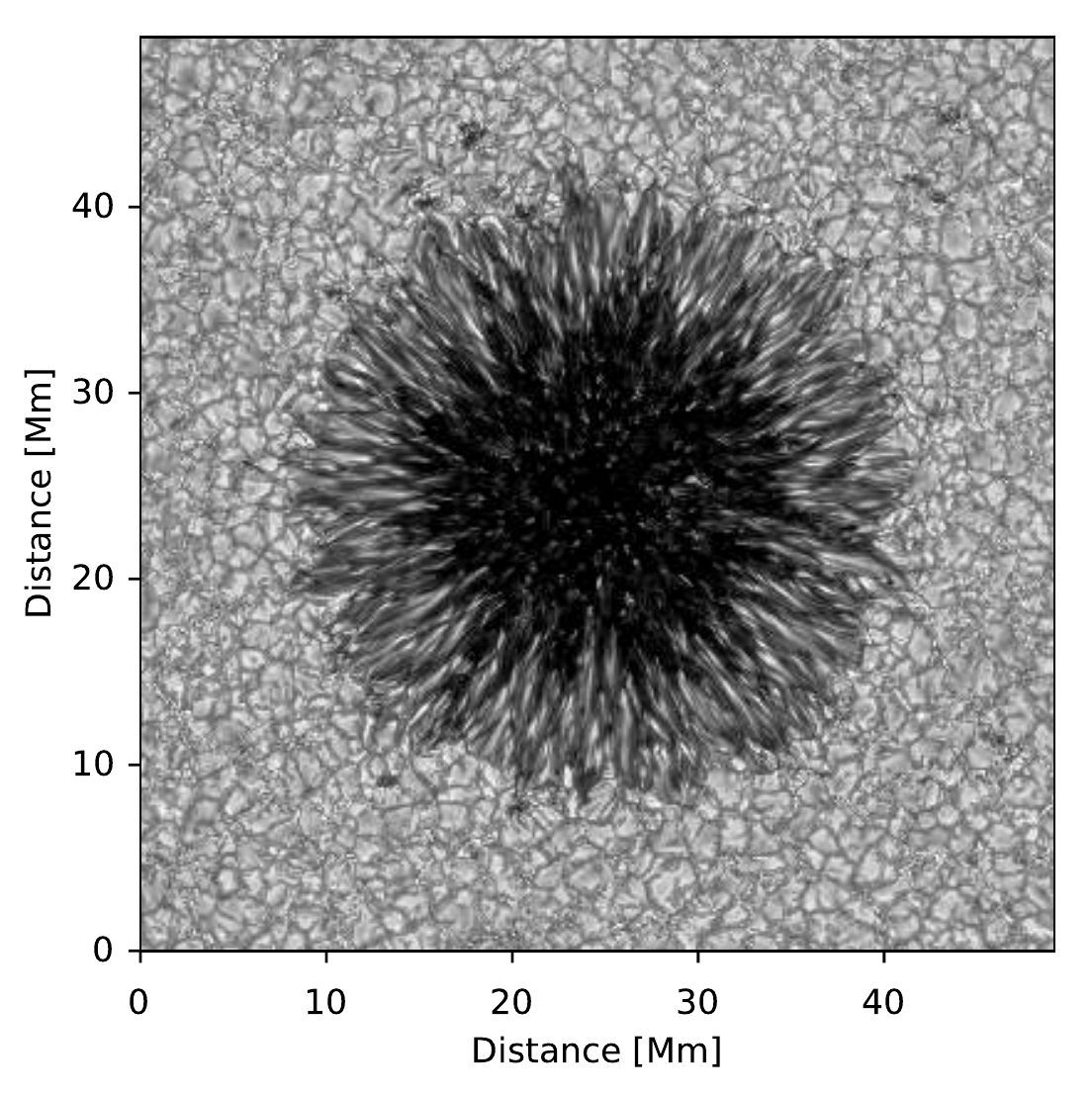}
    \includegraphics[width=0.6\textwidth]{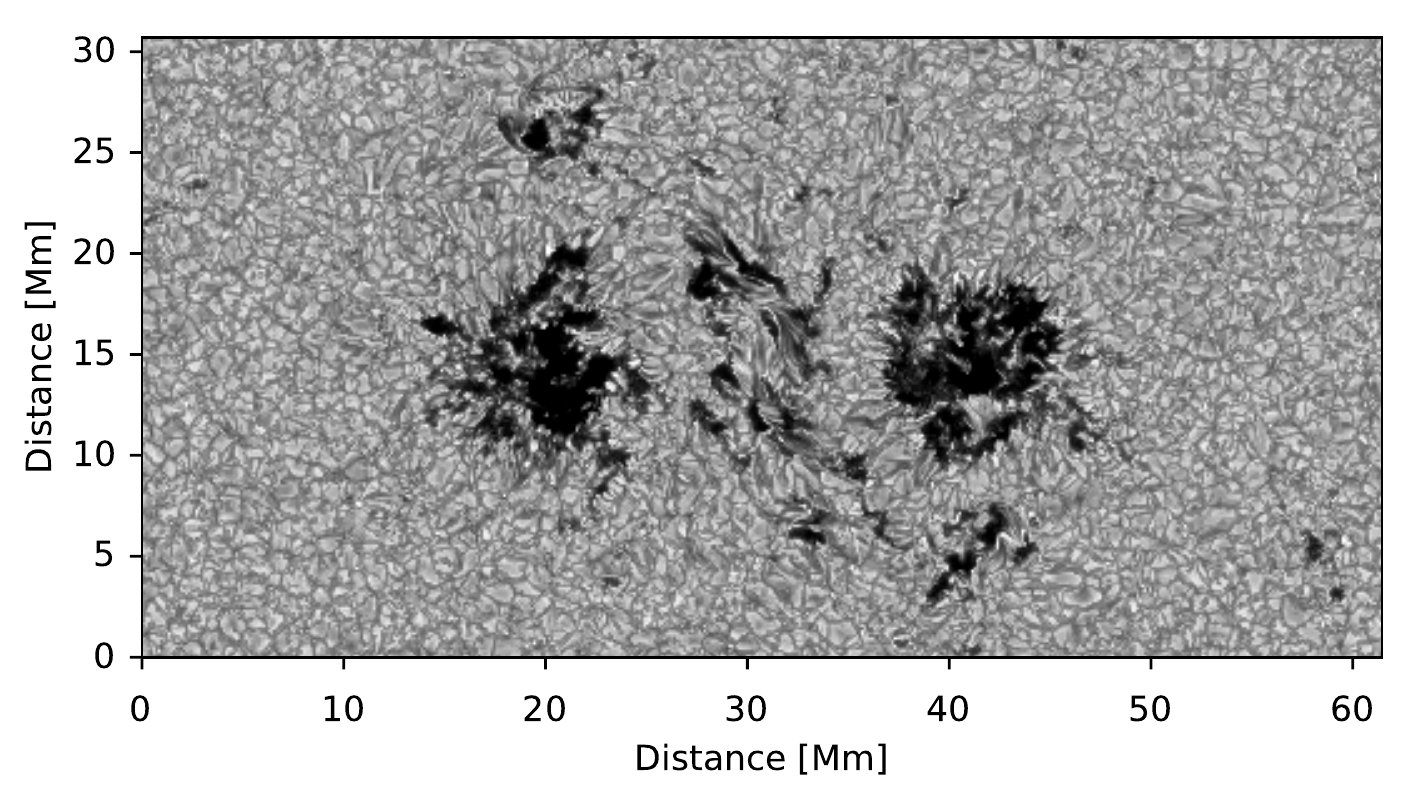}
    \caption{Synthetic continuum images of the snapshots used for training. The left 
    panel shows the sunspot simulation of \cite{Rempel12} and the right panel shows
    the emerging flux region simulation of \cite{Cheung2010}.}
    \label{fig:cont_rempel_cheung}
\end{figure*}

\section{Training sets}
Current DNNs are frequently limited only by the properties and quality of the
training set (apart from the availability of specialized hardware and time). 
For our purpose, two possibilities for training could be explored. The first one, that
we defer for a subsequent work, is to use standard inversions carried out with
codes like SIR. The 2D maps of inferred quantities are then used as training
for a deep convolutional neural network. One of the advantages of this approach
is that solar physicists are already familiar with these empirical inversions.
On the contrary, since the training is done in pixel-by-pixel inversions, the
use of CNNs during training only helps in denoising inversions but does not
exploit any spatial correlation present in the training data. The second option,
which is pursued here, is to use Stokes spectra synthesized in 3D MHD 
simulations of the solar atmosphere. These simulations might still suffer from a
lack of realism but they will eventually become very precise. The advantages of
training in these simulations are: i) the CNNs will be able to exploit all the
spatial information that is encoded in the training set, ii) deconvolution can
be easily implemented during training, iii) information that cannot be obtained using
standard inversions can be recovered (like the Wilson depression or the gas pressure).  
The disadvantage
of using simulations as training data is that results might be inaccurate. We 
expect the results to improve as more advanced simulations are developed.
In this sense, our results might be considered as a baseline that are
hopefully improved in the near future.

For our purpose we used two different snapshots carried out with the MURaM code \citep{vogler05}. 
The aim of this selection is to provide the neural network with examples of structures that
can be found later in real observation so that the properties
can be inferred with sufficient generalization. The first
one is a snapshot of the sunspot simulation carried out by \cite{Rempel12}. A continuum
image at 630~nm is shown in the left panel of Fig.~\ref{fig:cont_rempel_cheung}. 
This snapshot shows a well-developed sunspot, with
a penumbra of sufficient realism so as to generate the typical penumbral
filaments. The snapshot cube is of size 1536$\times$1536$\times$128 with 32~km 
horizontal (0.044\arcsec per pixel) and 16~km vertical grid spacing. This simulation is considered to be 
the state of the art in the generation of a numerical solar sunspot and its 
fine structure \citep{Tiwari2013}. One of the problems
of this simulation for our purpose is that the sunspot is unipolar, so that
a large percentage of the FOV contains only one polarity. For this reason, we 
artificially generate another snapshot where the magnetic field vector is reversed at each
individual grid point. This is a valid procedure under the assumption that this change does not
strongly affect the thermodynamics. We expect this to be the case
given that the photosphere is a region where the plasma-$\beta$ is large, so
the magnetic pressure is negligible in comparison to the gas pressure. We expect
in the future to improve our training set by adding another snapshot of a 
well-developed sunspot simulation with the opposite polarity.

The second snapshot is from the simulation of the formation of
an active region on the solar surface by \cite{Cheung2010}. The right
panel of Fig.~\ref{fig:cont_rempel_cheung}
displays the image in the continuum at 630~nm. The snapshot provides
a region of large complexity, with opposite polarities in the interface between the
two polarities of the active region. The snapshot is of size 1920$\times$960$\times$256
with 48~km (0.066\arcsec per pixel) horizontal and 32~km vertical grid spacing.

The synthetic Stokes profiles used during training are obtained after
the following process:
\begin{enumerate}
    \item The synthesis module of SIR\footnote{A parallel synthesis code for the 
    synthesis of LTE lines based on SIR is available on \url{https://github.com/aasensio/3d_sir}.}
    is used to compute the Stokes parameters
    for all the pixels in the two snapshots on the pair of Fe \textsc{i} lines at
    6301.5 \AA\ and 6302.5 \AA\ at disk center. SIR requires the
    the logarithm of the optical depth at 500~nm ($\log \tau_{500}$) as input. 
    For consistency, we compute this axis
    using the background opacity package inside SIR and discard all depths
    for which $\log \tau_{500}>2$ because they do not affect
    the emergent profiles. More precise background opacity
    packages could have been used but we have verified that a mismatch between the
    opacity package for the generation of the optical depth axis and that of
    the synthesis might affect the emergent Stokes profiles significantly, with 
    differences that can be as high as 15-20\%. For this reason, we prefer to
    use SIR as a reference to allow for better comparison with previous results
    and standard SIR inversions.
    \item The synthetic observations are spatially degraded with the Hinode PSF
    as computed by \citep{Danilovic2010}. The results are rebinned to the Hinode
    pixel size of 0.16\arcsec.
    \item The spectra at each spatial point is convolved with the spectral
    PSF of Hinode and reinterpolated on the standard Hinode wavelength axis of 112
    wavelength points. This makes a total of 448 wavelength points for each pixel when
    all Stokes parameters are serialized. Additionally, 
    the Stokes parameters are normalized to the median continuum in the surrounding quiet Sun.
    \item Stokes $Q$, $U$ and $V$ can span several orders of magnitude, typically in
    the range between 0.1 and 10$^{-3}$ (limited by the noise in Hinode). For this
    reason, we apply a normalization so that the input to the neural
    networks are around unity. We follow a different strategy in each architecture so that
    we explain them in detail in the next section.
    \item To mimic real Hinode observations, Gaussian noise with a standard deviation of 
    $10^{-3}I_c$, with $I_c$ the continuum intensity, is added to all Stokes profiles during training. 

%    \item We finally introduce a global jitter in velocity to all Stokes parameters in each
%    patch during training. This allows the network to learn that a global velocity shift in
%    the Hinode observations can be neglected. After some calibration, we
%    found that $\pm 10$ km s$^{-1}$ (equivalently $\sim10$ pixels) leads to good results. 

\end{enumerate}

The physical conditions that will be used during training are treated 
with the following process:
\begin{enumerate}
    \item The physical conditions are the temperature, $T$, the vertical velocity, $v_z$,
    the logarithmic gas pressure, $\log P$, and the three cartesian 
    components of the magnetic field, $\mathbf{B}$. They are obtained from the
    simulation cubes at seven corrugated surfaces of constant optical depth at 500~nm: $\log \tau=0$, 
    $-0.5$, $-1$, $-1.5$, $-2$, $-2.5$ and $-3$. Additionally, the geometric height
    associated to each one of these surfaces (the Wilson depression for each
    pixel and optical depth value) is stored. We subtract the average height of the
    $\log \tau=0$ surface on the quietest region of the snapshot. 
    This makes a total of 49 two-dimensional
    maps. These maps are rebinned to the Hinode resolution
    but not degraded with any PSF. In fact, this is somehow problematic because
    it is well-known that the average Stokes profiles emerging from a set of
    atmospheres is not equal to the synthetic Stokes profiles in the average
    atmosphere \citep[e.g.,][]{Uitenbroek11}. However, the impact in our case
    is not too severe because we only need to 
    average 2.4$\times$2.4 pixels for Cheung's simulation and 3.6$\times$3.6 pixels
    in the case of Rempel's snapshot. Other options, that we defer for a future
    analysis, imply
    applying superresolution in the DNNs that we consider in Sect.~\ref{sec:architecture}
    and reduce the effect of the rebinning, eventually going up to the original resolution
    of the simulation. 
    \item One of the main issues for inferring the magnetic field is the presence
    of ambiguous solutions. In the Zeeman regime (the one that we assume for the synthesis of
    the Stokes profiles) we only find the 180$^\circ$ ambiguity in the azimuth on the
    LOS. Since we assume observations at disk center, this means that a magnetic field with an azimuth 
    $\phi$ and another one with an azimuth $\phi+\pi$ gives exactly the same
    Stokes parameters. If this ambiguity is left in the training set, the neural
    network cannot choose the right solution, so it is mandatory to remove it for
    a stable training. We do this by using the following three quantities as output instead of 
    the three cartesian components of the magnetic field $(B_x,B_y,B_z)$:
    \begin{align}
        B_Q &= \mathrm{sign}(B_x^2-B_y^2){|B_x^2-B_y^2|}^{1/2} \nonumber \\
        B_U &= \mathrm{sign}(B_x B_y){|B_x B_y|}^{1/2} \nonumber \\
        B_V &= B_z.
    \end{align}
    The motivation for the labels of the variables is that, in the case of constant
    magnetic field along the LOS, Stokes $Q$, $U$ and $V$ are proportional to $B_Q$, 
    $B_U$ and $B_V$, respectively. Their advantage is that they are not affected by 
    the 180$^\circ$ ambiguity. The cartesian components of the magnetic field
    can be obtained by carefully inverting the previous equations and selecting
    only one of the two compatible solutions.
\end{enumerate}

After all degradations and rebinning, the emerging flux simulation turns out to
be of size 397$\times$794 pixels, while
the sunspot simulation ends up being of size 424$\times$424 pixels. From these two simulations
we randomly extract 50,000 patches of 32$\times$32 pixels with equal spatial probability for
training and an extra dataset of 2,000 for validation. 
The difference in area occupied
by the quiet Sun, umbra and penumbra in the two snapshots is not large, so we find
unimportant to spatially bias the probability to equally cover the three types of
regions. Given that
we only have two snapshots, we heavily rely on augmenting the training set by randomly flipping
the patches horizontally and vertically and also by rotations of the patches by 90, 180 and 270$^\circ$.

The training of the neural networks is done by optimizing the following scalar loss function,
where the summation is carried out over all pixels of all patches in the training set:
\begin{equation}
\label{eq:loss}
    L = \sum_i \left\lVert \mathbf{T}_i - \mathbf{f}(\mathbf{W};\mathbf{S}_i) \right\rVert^2.
\end{equation}
The loss function measures the $\ell_2$ distance between the $i$-th 
target patch with the physical conditions, $\mathbf{T}_i$,
and the output of the neural network with weights $\mathbf{W}$ and input Stokes parameter 
for the very same $i$-th patch, $\mathbf{f}(\mathbf{W};\mathbf{S}_i)$. Architecture-dependent
details of the training are explained later.

\section{Architectures}
\label{sec:architecture}
We describe in the following the two architectures that we have used, pointing out 
their main computational advantages and disadvantages. Contrary to previous 
studies \citep{Carroll2001,Carroll2008} in which 1D neural networks were trained, 
we have used CNNs to exploit the 2D
spatial coherence of the FOV. This strategy strongly decreases the
model degeneracy due to local noise or multiple minima and also allows
us to invert the whole map in one single application of the neural networks.

\begin{figure*}
    \centering
    \includegraphics[width=\textwidth]{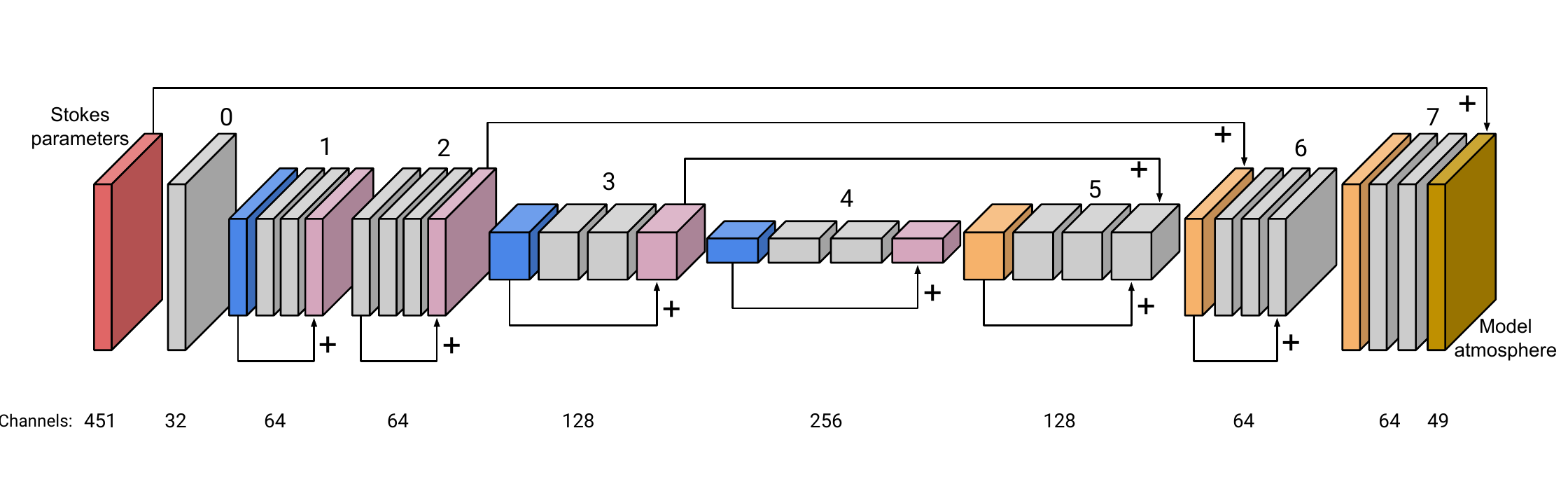}
    \includegraphics[width=0.9\textwidth]{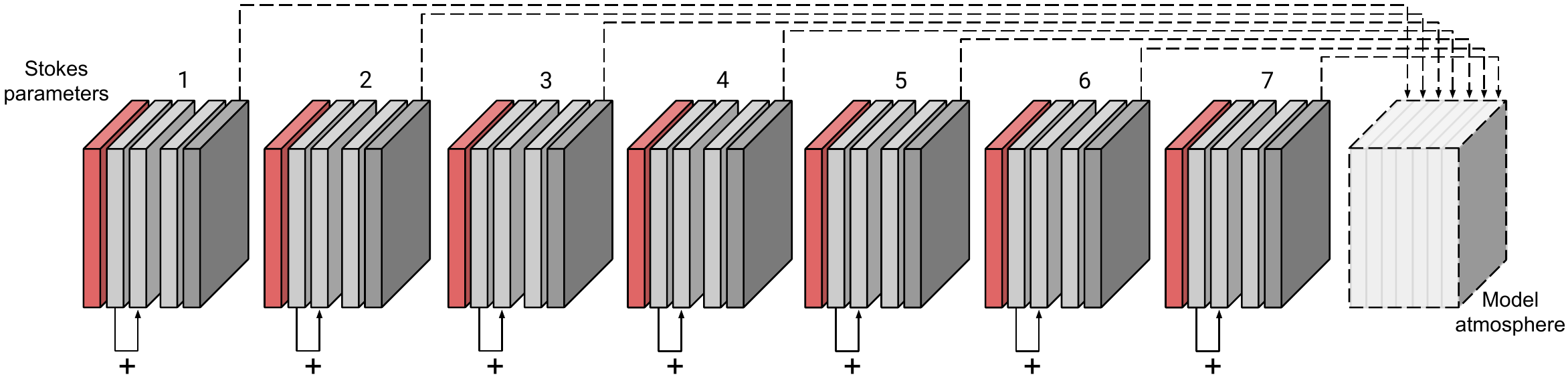}
    \caption{Encoder-decoder (upper panel) and concatenate architectures (lower panel).}
    \label{fig:encdec}
\end{figure*}

\subsection{Encoder-decoder}
The first architecture, displayed in the upper panel of Fig.~\ref{fig:encdec} follows the 
fully convolutional encoder-decoder paradigm, in which the input data is
first analyzed by an encoder neural network that acts as a funnel by reducing the spatial
dimensionality of the images while increasing the number of channels. This induces the
neural network to extract relevant spatial information and encode it in the channel
dimension. After the encoding phase, a decoder neural network recovers the original size
of the input images. In this case, the only restriction is that, if one wants 
to get an output of the same dimension as the input, then the input size has to 
be an integer multiple of $2^3$. If this is not the case, the output will 
have a slightly different number of pixel than the input.

We previously used this architecture in \cite{DeepMFBD18} with fantastic
performance. Both the encoder and the decoder are based on the concept of residual
networks \citep{residual_network16}, in which skip connections directly connect some layers with later
layers. Recent analysis suggest that this not only accelerates training, but also
very efficiently removes local minima from the loss hypersurface \citep{Du2018}. We utilize
skip connections inside each set of blocks (shown as arrows in the lower part of the
blocks) and also between blocks of the same spatial size (shown as arrows in
the upper part of the blocks). Additionally, we use the fact that some output parameters
have some resemblance with continuum images. This is the case for
the temperature and Wilson depression. For these outputs, we simply add
the continuum image (in normalized units) directly to the output of the decoder.

The specificities of the encoder-decoder architecture are as follows.
To normalize the polarization and prevent it from being ignored by being
much smaller than Stokes $I$, we compute the map of maximum amplitude
in absolute value (thresholded from below at 5$\times$10$^{-3}$) for 
each Stokes parameter $Q$, $U$ and $V$ and normalize
them to each value\footnote{We have verified that, when run in validation
mode with real Hinode data, slightly better results are found if the threshold is 
set to a larger value (of the order of $0.1$), with no appreciable impact
on the inferred quantities.}. 
The maps of maximum amplitude are then added as input
in log scale after subtracting the mean value. Therefore, the total number of input 
channels is 448+3=451. For that reason, 
the red blocks in Fig.~\ref{fig:encdec}, which refer to input layers are, 
in this case, patches of size 32$\times$32
with 451 channels. Concerning the output, all physical conditions are linearly transformed
to the interval $[0,1]$ using the minimum and maximum values from all available
snapshots. When the neural
network is used in evaluation, this transformation is undone to recover 
all physical quantities in the correct units.
Finally, we point out that we add a noise realization to the 
maps of the snapshots used for training and leave it unchanged during the training.

In the encoder-decoder architecture, gray blocks refer to convolutional blocks. They are made 
of the consecutive application
of a batch normalization layer \citep{batchnormalization15}, a rectified linear unit activation
function \citep[ReLU;][]{relu10} and a convolution with a kernel of size 3$\times$3 with
a reflection padding of one pixel in the borders to maintain the size of the images. Pink
blocks are similar to the gray ones but the convolution is done with a 1$\times$1 kernel. Blue
colors are similar convolutional blocks but the convolution is done with a stride of 2, thus
reducing the size of the input image by a factor 2, but
increasing the number of channels by another factor 2. Orange blocks are made of the consecutive
application of batch normalization, ReLU, a bilinear upsampling of a factor 2 and
a 3$\times$3 convolution with reflection padding while reducing the number of 
channels. Finally, the golden layer is the 
output, in our case patches of size 32$\times$32 with 49 channels. The number of channels
of each set of blocks is indicated in the figure. 

We note that the specific number of blocks
and their properties have been set according to previous experience and also
encouraged by the successful results showed in the next sections. Although not
proven here, we assume that a careful ablation study can produce simpler and 
faster architectures. Specifically, our election of 32 channels
in the first convolutional block is motivated by the expected dimensionality of
spectropolarimetric data in the pair of Fe \textsc{i} lines at 630 nm \citep{2007ApJ...660.1690A}
and the fact that even a simple linear principal component analysis of the Stokes 
parameters is able to reconstruct the profiles close to the noise level with only 
a few eigenprofiles.

\begin{figure*}
    \centering
    \includegraphics[width=0.70\textwidth]{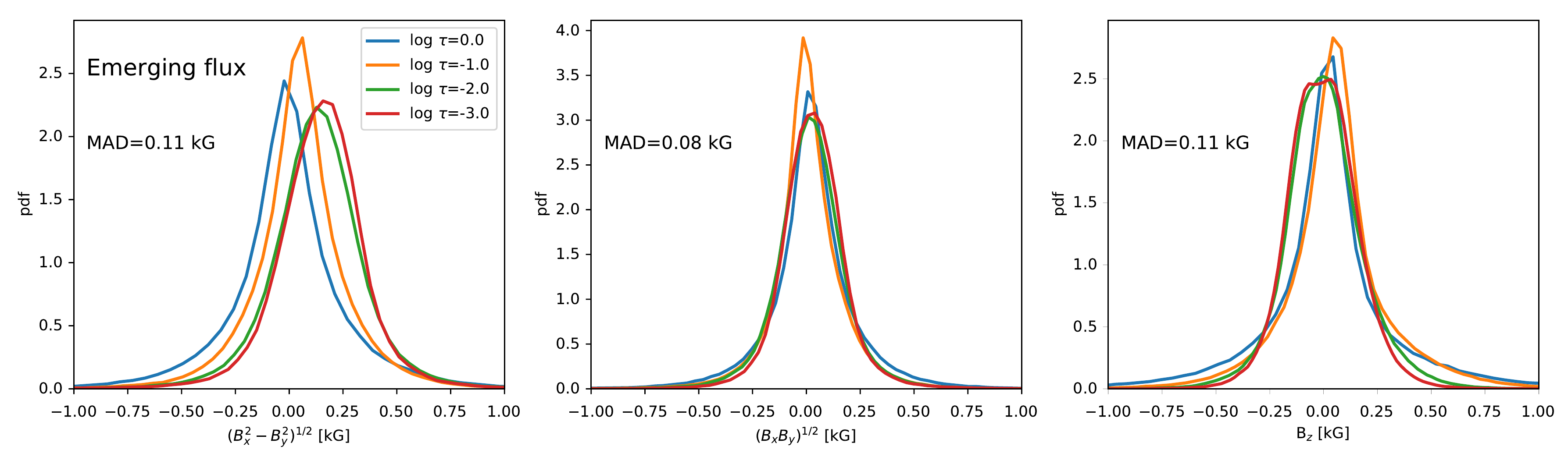}
    \includegraphics[width=\columnwidth]{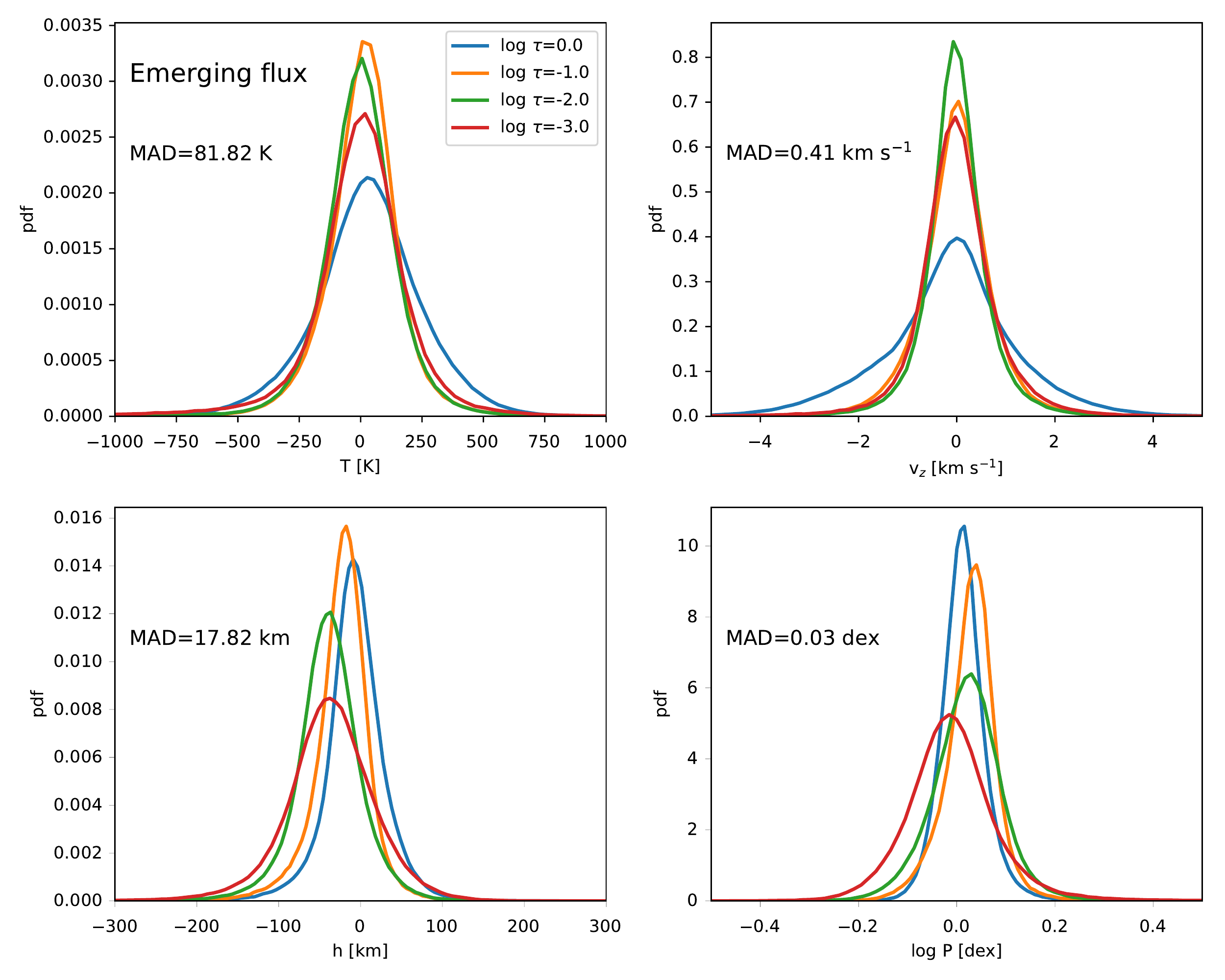}
    \caption{Gaussian kernel density estimate of the difference between the physical parameters
    at different constant optical depth surfaces and the value inferred by the encoder-decoder DNN
    in the validation set. Results are similar for both architectures.}
    \label{fig:histograms_validation}
\end{figure*}

One of the advantages of the encoder-decoder architecture is that spatial information
is efficiently shared. After the three stages of encoding that we use, an
image that is originally of size 32$\times$32 turns out to be of size 4$\times$4. A convolution with
a kernel of size 3$\times$3 thus couples all pixels in the image at this fourth stage. 
The output of the network 
can therefore make efficient use of spatial information from distant 
structures in the Stokes parameters.
%This makes
%that the output of the network can make efficient use of spatial information from structures
%in the Stokes parameters that were very distant apart. 
This is probably not crucial
in our case because Stokes parameters in LTE fundamentally depend on very local 
information\footnote{For the 3D non-LTE problem 
this type of neural networks could be very efficient on capturing the relation 
between each pixel and its surroundings.}
(not much larger than the size of the PSF). However, it introduces an extra robustness
that can be very interesting for very extended structures like large sunspots.
From a computational perspective, encoder-decoder architectures 
are memory- and computation-efficient. Images become smaller
during the encoder phase, so that convolutions can be made faster and
with a lower memory footprint. On the disadvantage side, encoder-decoder
networks are sometimes slightly difficult to train.

The scalar loss of Eq.~(\ref{eq:loss}) is optimized with respect
to $\mathbf{W}$ via the well-known Adam optimizer \citep{adam14}.
The network has $\sim$3.31 million free parameters. 
The neural network was trained during 50 epochs with a batch size of 128. 
The learning rate is 3$\times$10$^{-4}$ but it is reduced by a factor of $1/2$ every 
30 epochs. 
The neural network is implemented and trained in \texttt{PyTorch}, which
seamlessly allows us to leverage Graphical Processing Units (GPU) to accelerate
the calculations \citep{Harker2012}. We used a Titan X and a P100 NVIDIA GPUs for training
during the exploration of hyperparameters.

\subsection{Concatenate}
The second architecture, displayed in the lower panel of Fig.~\ref{fig:encdec}, 
is also fully convolutional. The main difference with the encoder-decoder architecture
is that the spatial size of the images is kept fixed throughout the entire architecture. 
For this second architecture we have opted for a more conservative strategy, similar in
philosophy to \cite{Carroll2001} and \cite{Carroll2008}, who trained a different network for each physical parameter. 
The main difference is that the output of mini-networks are concatenated at the end of the 
topology and we include more physical parameters than previous studies.

One of the advantages of this architecture is that the size in pixels of the input can be 
arbitrary and does not need to be multiple of any number. While in the encoder-decoder 
network the spatial dimensionality of the input is reduced thanks to the bottleneck, a 
similar compression effect can be achieved in this case by reducing the flexibility 
of the connectivity or reducing the number of kernels of each convolutional layer
in the network. On the downside, by having a mini-network for each physical 
variable, this architecture does not exploit
the common patterns and relationships among the different physical parameters.
%\footnote{We point 
%out that precisely these common patterns can sometimes induce
%pathological behaviors on DNNs. For instance, we noted that in some cases the velocity field was
%extracted purely from the spatial behavior due to its similarity with the
%temperature map, almost neglecting the wavelength shift of the
%Stokes profiles. This pathology was fixed by adding the velocity jitter 
%during training in the encoder-decoder network.}. 
Consequently, it needs a large number of network 
parameters to achieve high accuracy. Despite these differences, many of the 
details discussed in the previous section about accelerating the training
and improving the accuracy still applies to this case.

The specificities of the concatenate architecture are as follows. As before, 
the red block in the lower panel of Fig.~\ref{fig:encdec} refers to input layers, 
in our case patches of size 32$\times$32 with 448 channels, that is, just
four Stokes parameters with 112 wavelength points each. The normalization in
this case is very simple (1 for Stokes $I$ and 0.1 for Stokes $Q$, $U$ and $V$). 
%We have proven that 
The accuracy of the network is good using this simple scaling, which
makes all Stokes parameters have roughly the same order of magnitude.
Finally, noise is added online during the creation of each batch, so it 
changes from epoch to epoch.

Gray blocks in the lower panel of Fig.~\ref{fig:encdec} refer to 
convolutional blocks, which are slightly different to those of the
encoder-decoder case. They are built by the consecutive application 
of a convolution with a kernel of size 3$\times$3 with
a reflection padding and an exponential linear 
unit activation function \citep[ELU;][]{ELU2015}. This activation function
has been proven to speed up learning while leading to higher accuracies than 
using batch normalization layer followed by a ReLU activation
function, like in the encoder-decoder architecture. ELUs allow negative values
to pass through the network, thus forcing the mean response of the neurons
to lie closer to zero but with a clear 
saturation plateau in its negative regime. This allows networks to learn 
more robust and stable representations. Additionally, the loss function
presents a smoother decrease during convergence as compared with using ReLUs.

\begin{figure}[!ht]
    \centering
    \includegraphics[width=0.9\columnwidth]{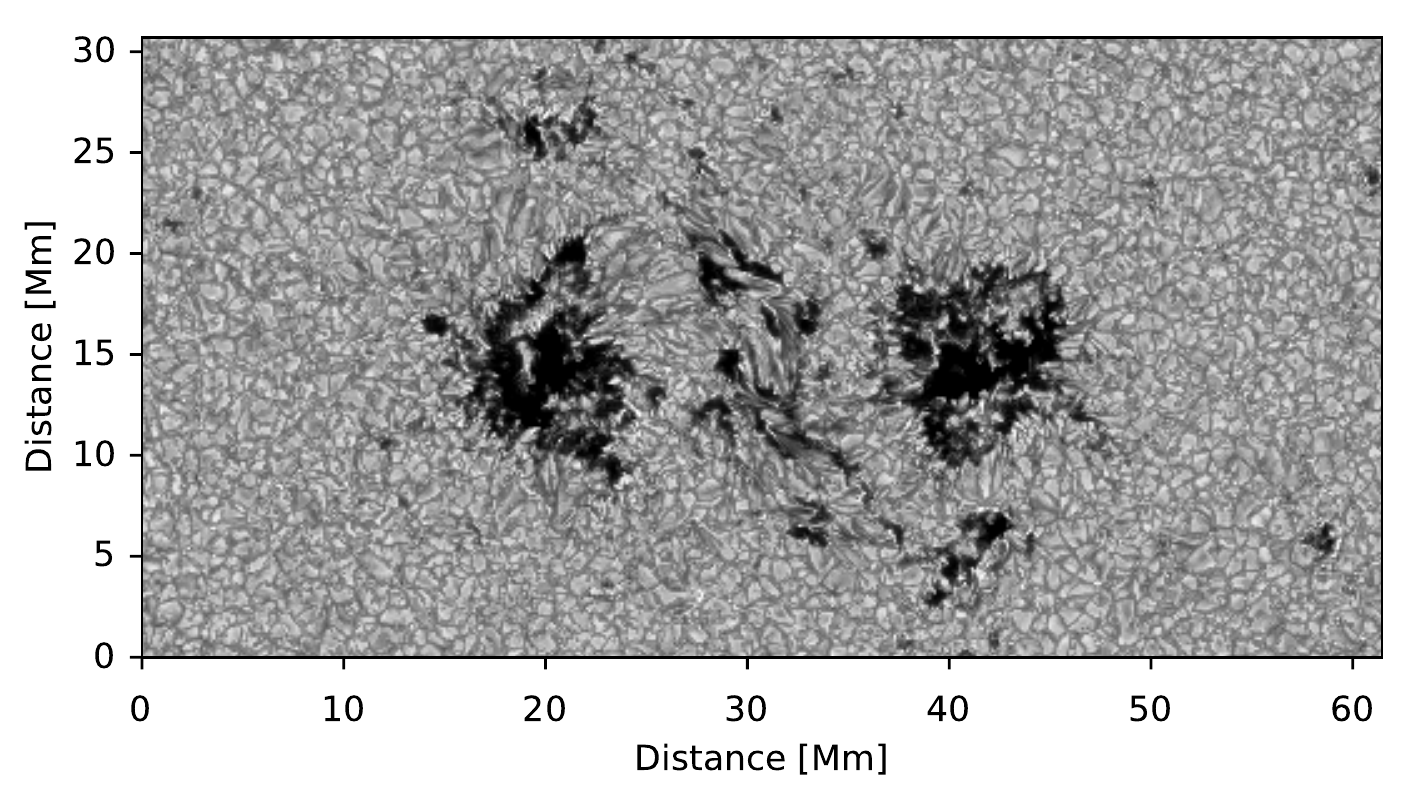}
    \caption{Synthetic continuum image of the validation snapshot 
    from the emerging flux region simulation of \cite{Cheung2010}.}
    \label{fig:cont_validation}
\end{figure}

While the first two gray blocks have 32 channels each, the last gray blocks 
has 7 channels to match the number of depths used for each parameter. The last 
dark gray block is the only block that performs convolutions without activation.
The result of each dark gray block is then concatenated at the end
for producing the output without any additional transformation (indicated
with dashed arrows above the network).
Similarly to the previous case, each physical parameter is transformed to be close 
to unity when possible by using the following units: the temperature in kK, the velocity in km~s$^{-1}$, 
the magnetic field components in kG, and the geometrical height 
in Mm\footnote{We point out that the differences in the normalization between
the two architectures are purely due to the blind designing process of both architectures.}.
Since the spatial size of the images is kept in each convolutional layer,
we do not utilize any pooling or upsampling operation. 
Each output of the concatenate architecture is the result of 4 convolutional blocks with
$3 \times 3$ kernels. Although the first block produces a receptive field 
of size $3 \times 3$ for each neuron, the second block increases that to $5 \times 5$, the
third one to $7 \times 7$ and the last one to $9 \times 9$. Such a receptive field
is able to couple, for a single pixel in the output, the Stokes parameters of
a patch of $\sim 1.44''$, which is almost 5 times the diffraction limit of 
Hinode/SOT. This size is very similar to the size of the PSF considered in the
coupled inversions carried out by \cite{vannoort12}.

The total number
of free parameters in this architecture is $\sim 1$~M free parameters. We train the network
using Adam with a constant learning rate of $10^{-4}$ during 20 epochs 
with a batch size of 10. This neural network was implemented and trained in \texttt{Keras}, which
also allows us to leverage GPUs to accelerate the calculations.

\begin{figure*}
    \centering
    \includegraphics[width=\textwidth]{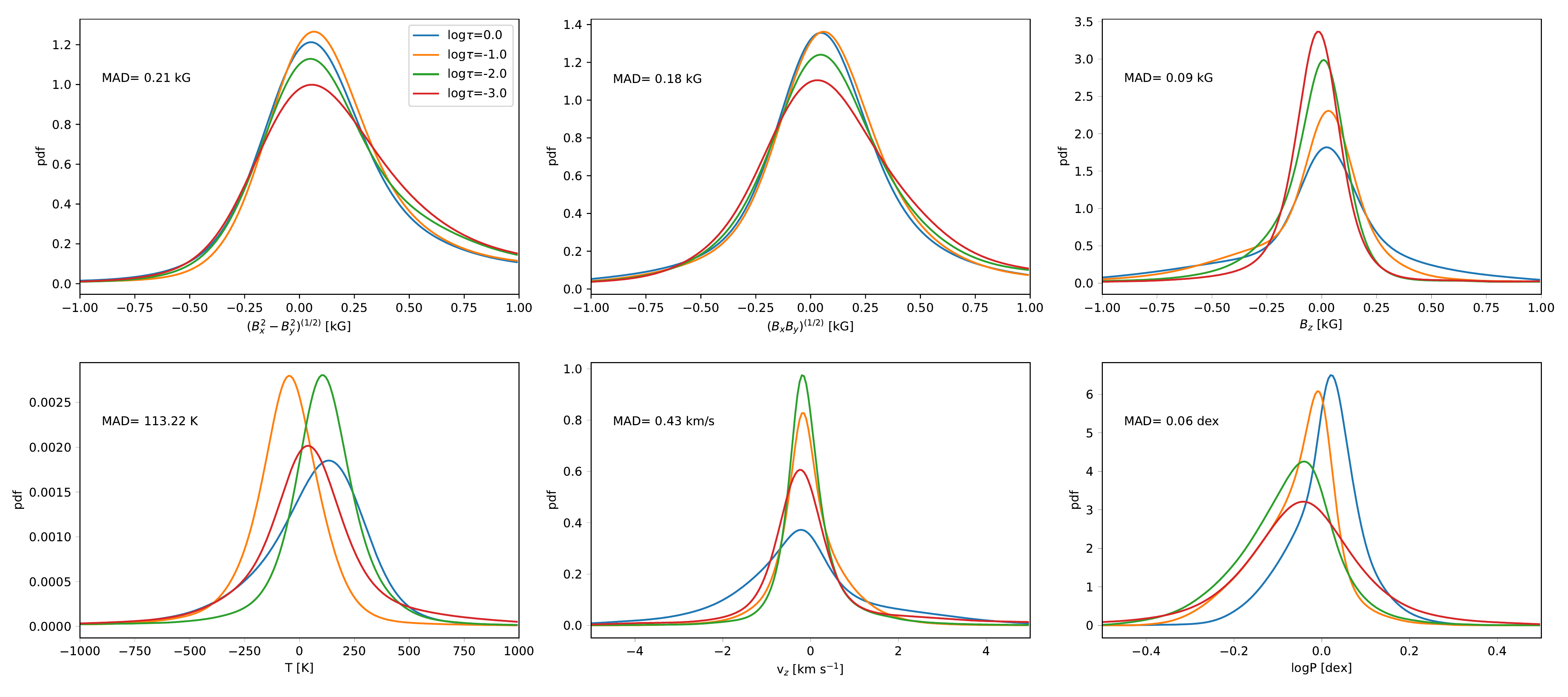}
    \caption{Gaussian kernel density estimate of the difference between the physical parameters
    at different constant optical depth surfaces and the value inferred with SIR.}
    \label{fig:histoSIR}
\end{figure*}

\section{Results}

Once the DNN is trained, it can be applied to any Hinode observation, provided
that the data is preprocessed in the same way. This preprocessing requires simply
to normalize to the local average quiet Sun and carry out the specific
normalizion for Stokes $Q$, $U$ and $V$. Given the fully convolutional character
of the trained networks, they can be applied to any input, irrespectively of
their dimension (taking into account the caveats of the encoder-decoder network). 
One of the huge advantages
of neural networks is their enormous speed during inference. As an example, in a
low-profile GPU like the NVIDIA Quadro M4000, the inference of a Hinode map of 
512$\times$512 pixels can be done in $\sim$180 ms, which turns out negligible
if compared with the time spent in any input/output operation. This speed
can be much higher if several maps are used as input as part of
a batch, which then would take better advantage of the parallel computation of
GPUs. Obviously, using a more modern GPU  
can cut these times by another appreciable factor. As a consequence, assuming
zero overhead in input/output operations, one can carry out 480,000 inversions of 512$\times$512
maps per day with this neural network. The current Hinode database for SOT/SP contains of the
order of 12$\times$10$^3$ groups of observations. Assuming that all these groups
refer to maps of the quoted size (which is a very conservative upper limit), one can
carry out the inversion of the whole Hinode SOT/SP with our proposed neural network
in $\sim$36 minutes.

\begin{figure*}[!ht]
    \centering
    \includegraphics[width=0.98\columnwidth]{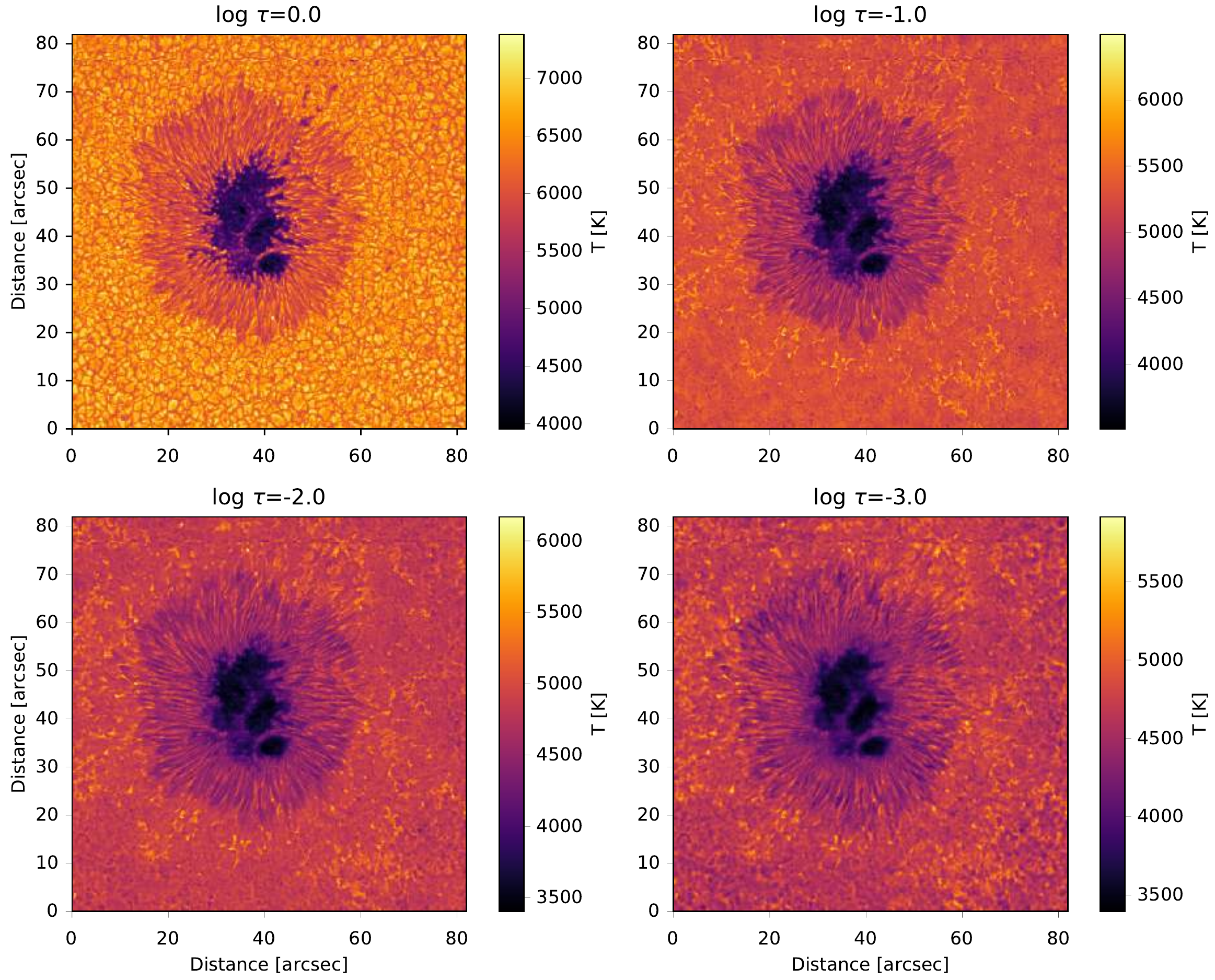}%
    \includegraphics[width=0.98\columnwidth]{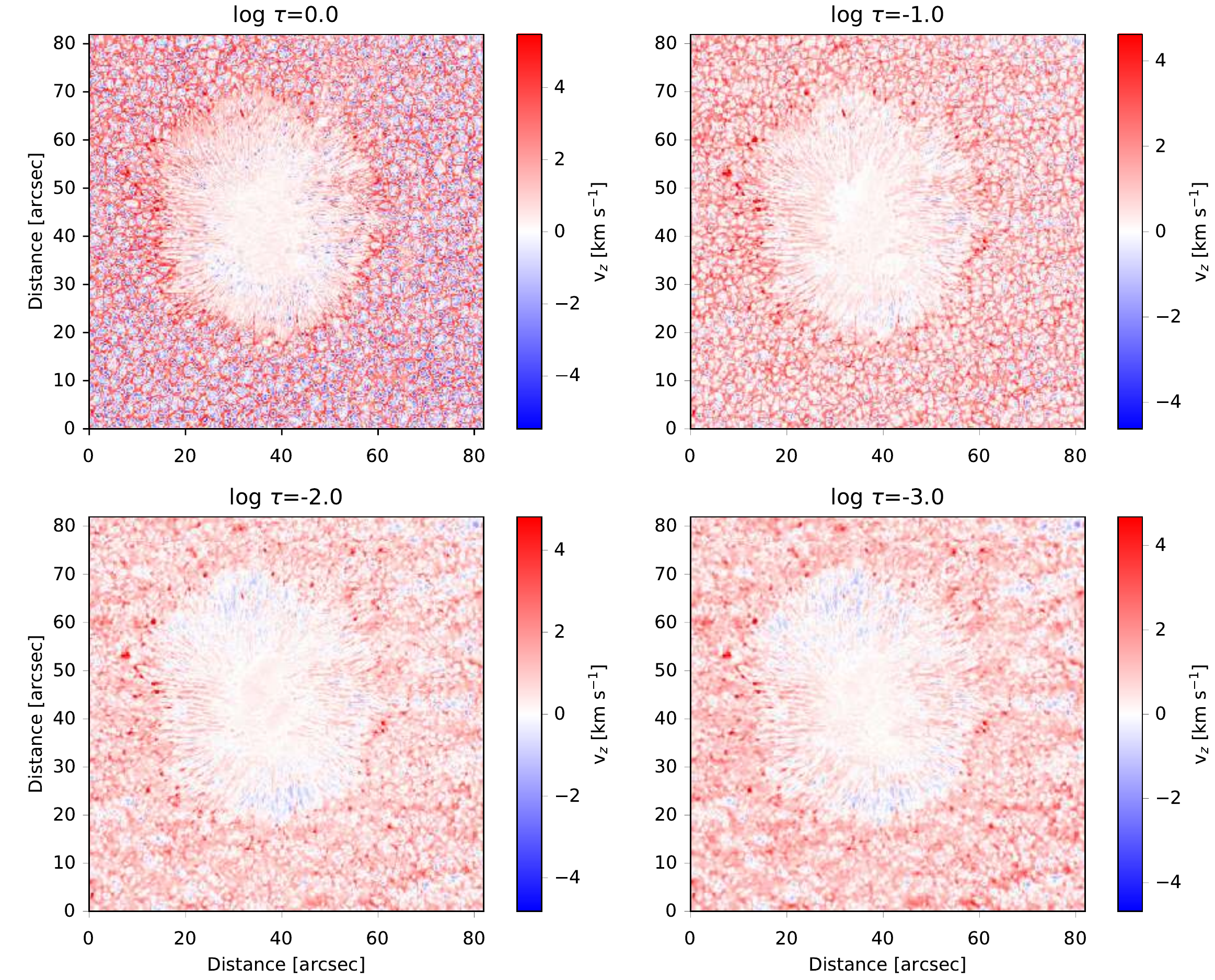}
    \includegraphics[width=0.98\columnwidth]{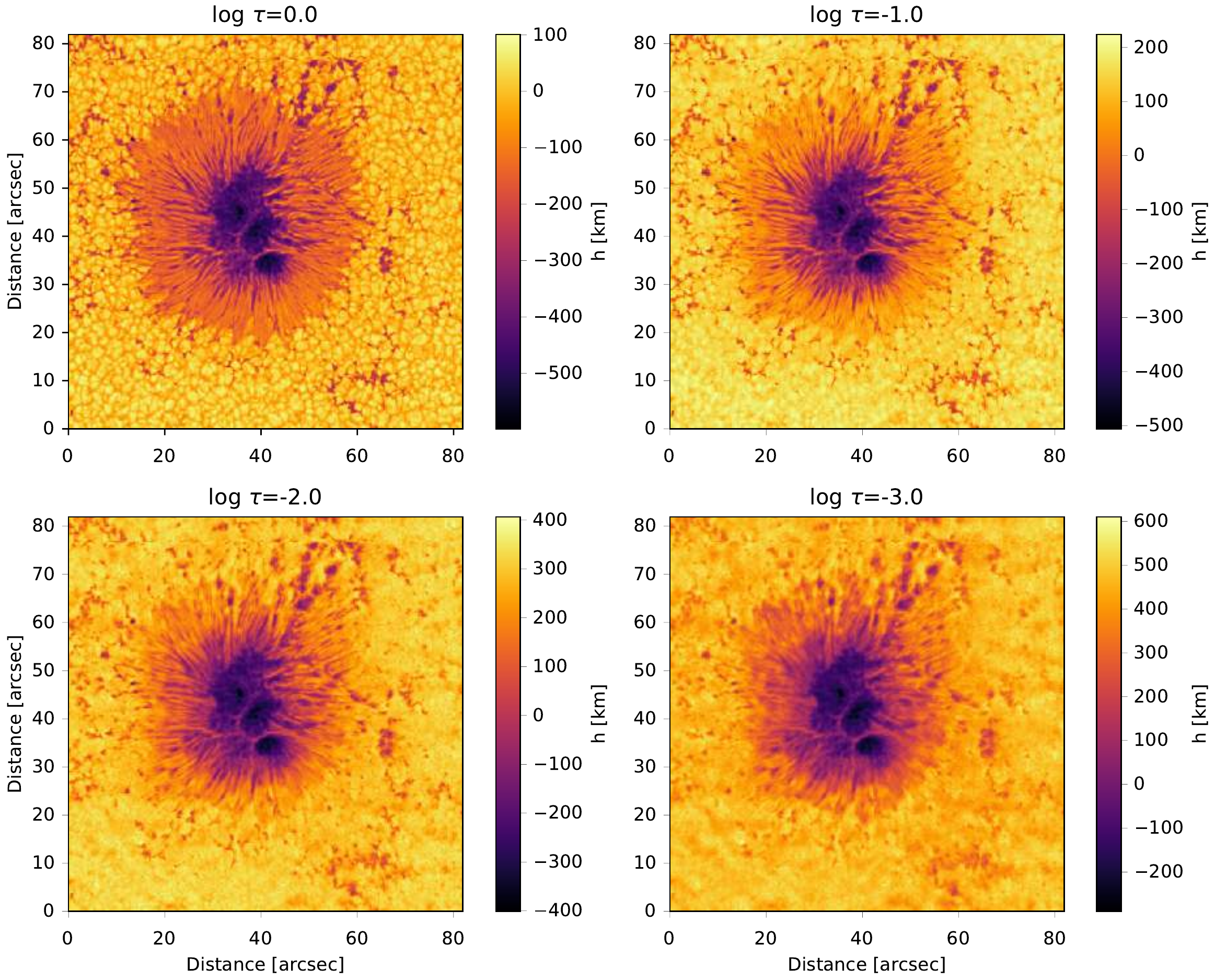}%
    \includegraphics[width=0.98\columnwidth]{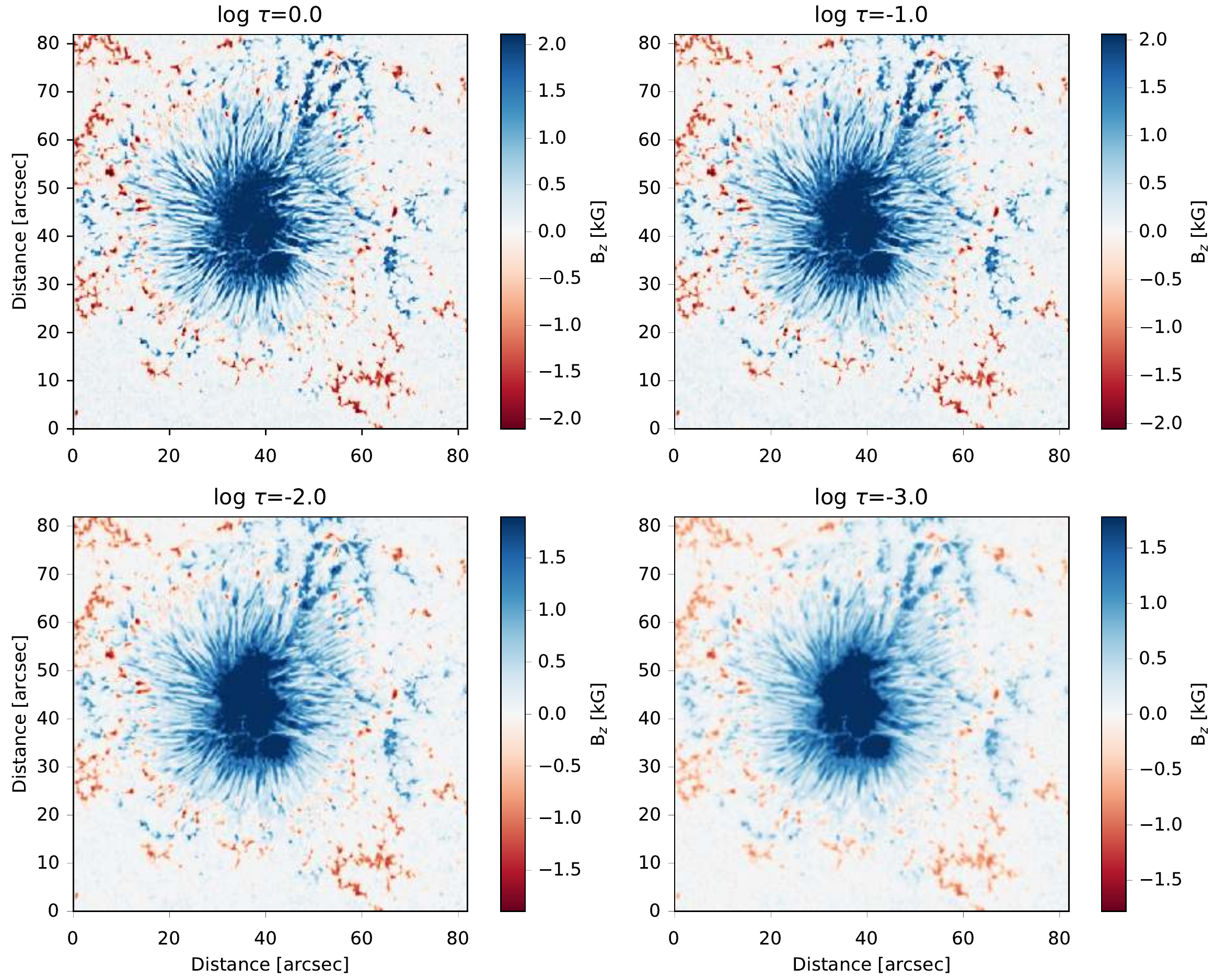}
    \includegraphics[width=0.98\columnwidth]{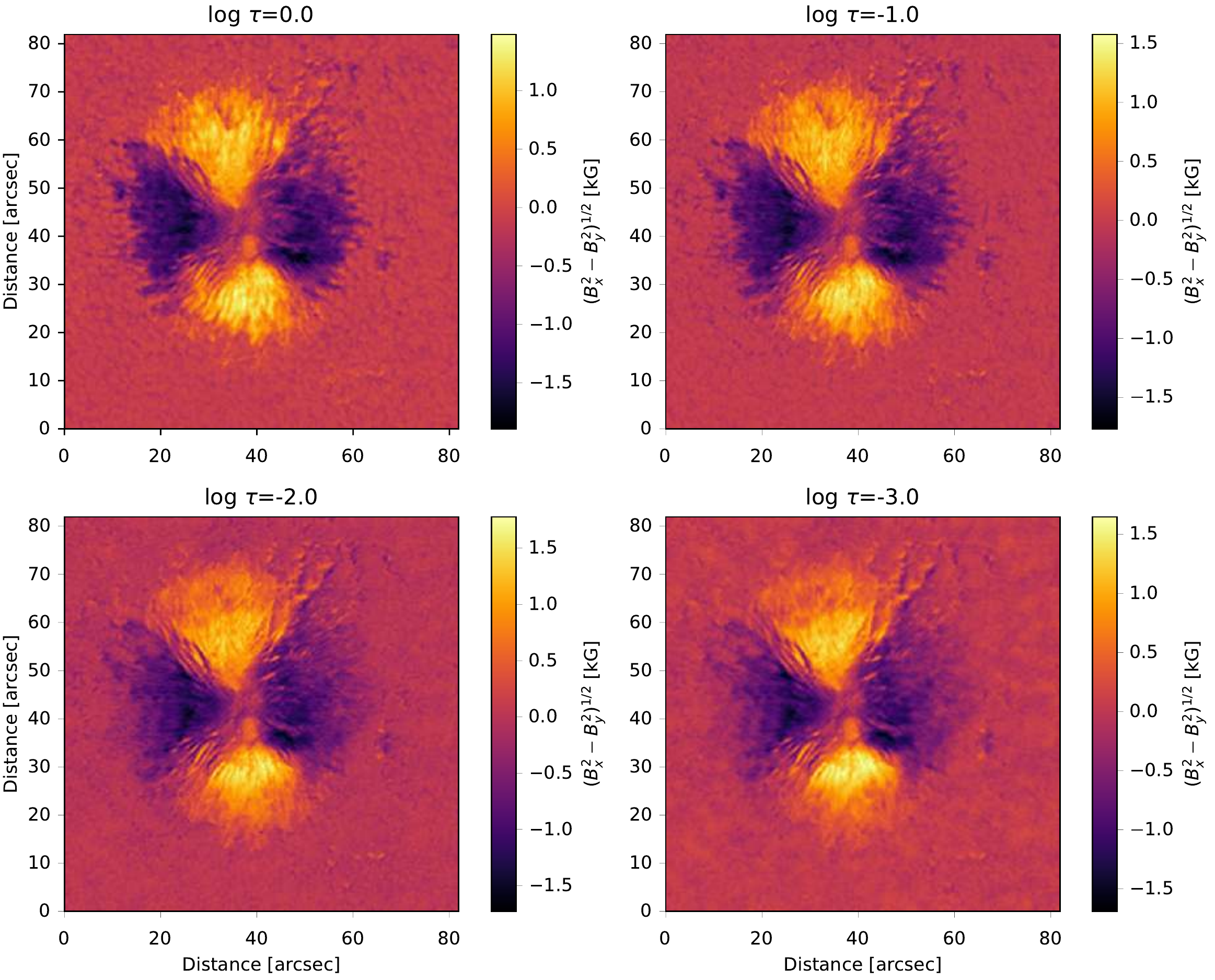}%
    \includegraphics[width=0.98\columnwidth]{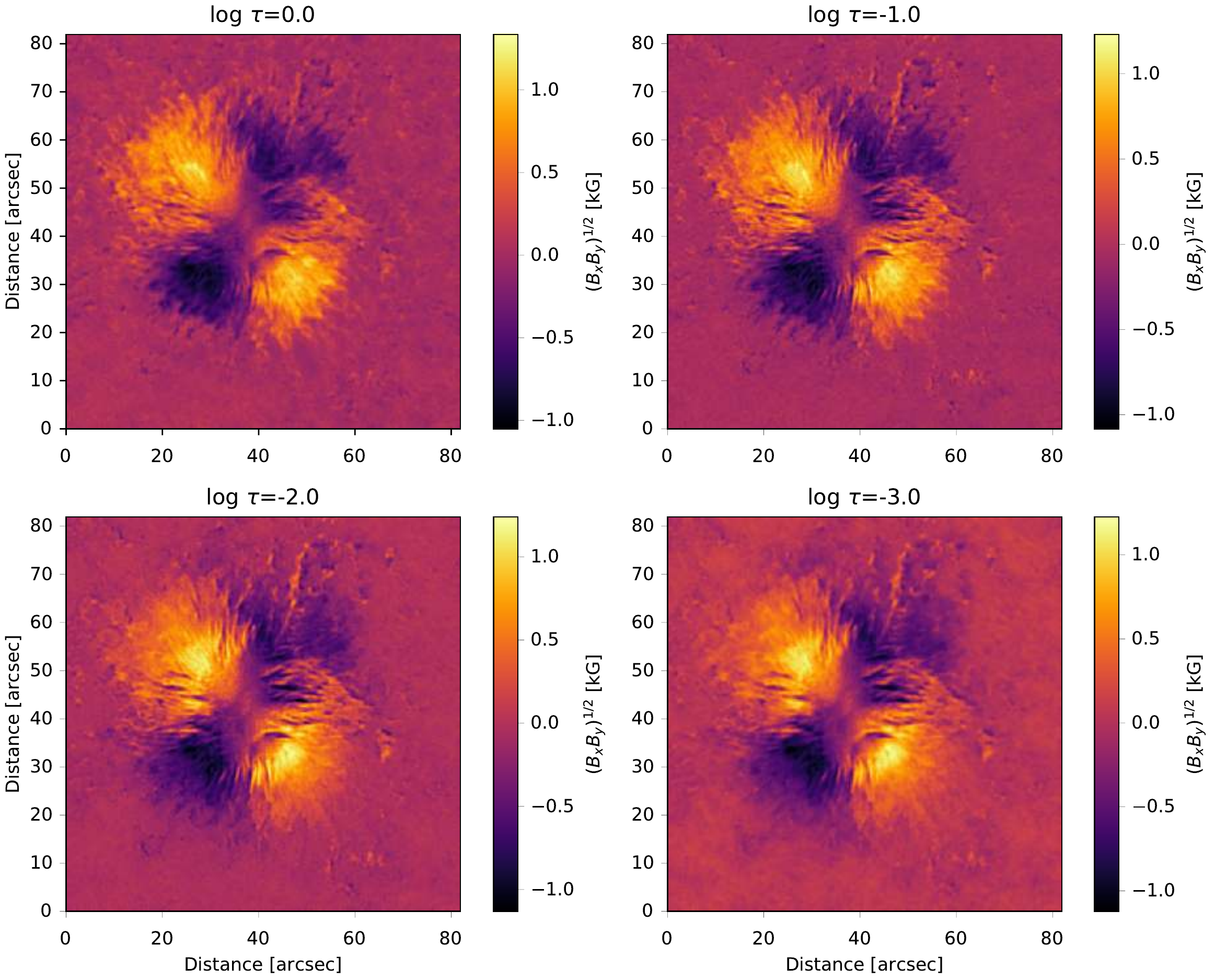}
    \caption{Predicted physical properties of AR10933 using the encoder-decoder network.
    \label{fig:maps_ar10933}}
\end{figure*}

\begin{figure*}[!ht]
    \centering
    \includegraphics[width=0.98\columnwidth]{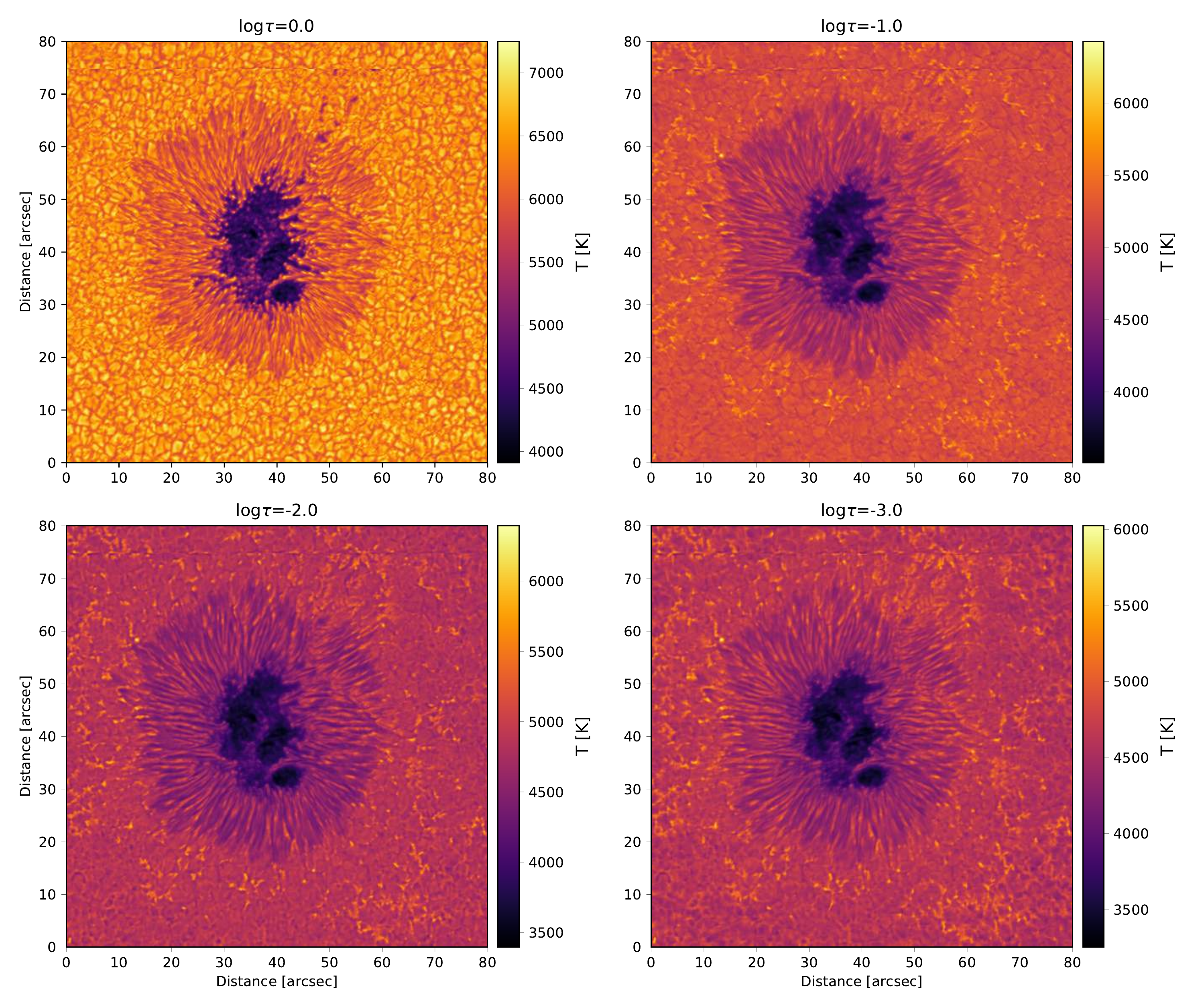}
    \includegraphics[width=0.98\columnwidth]{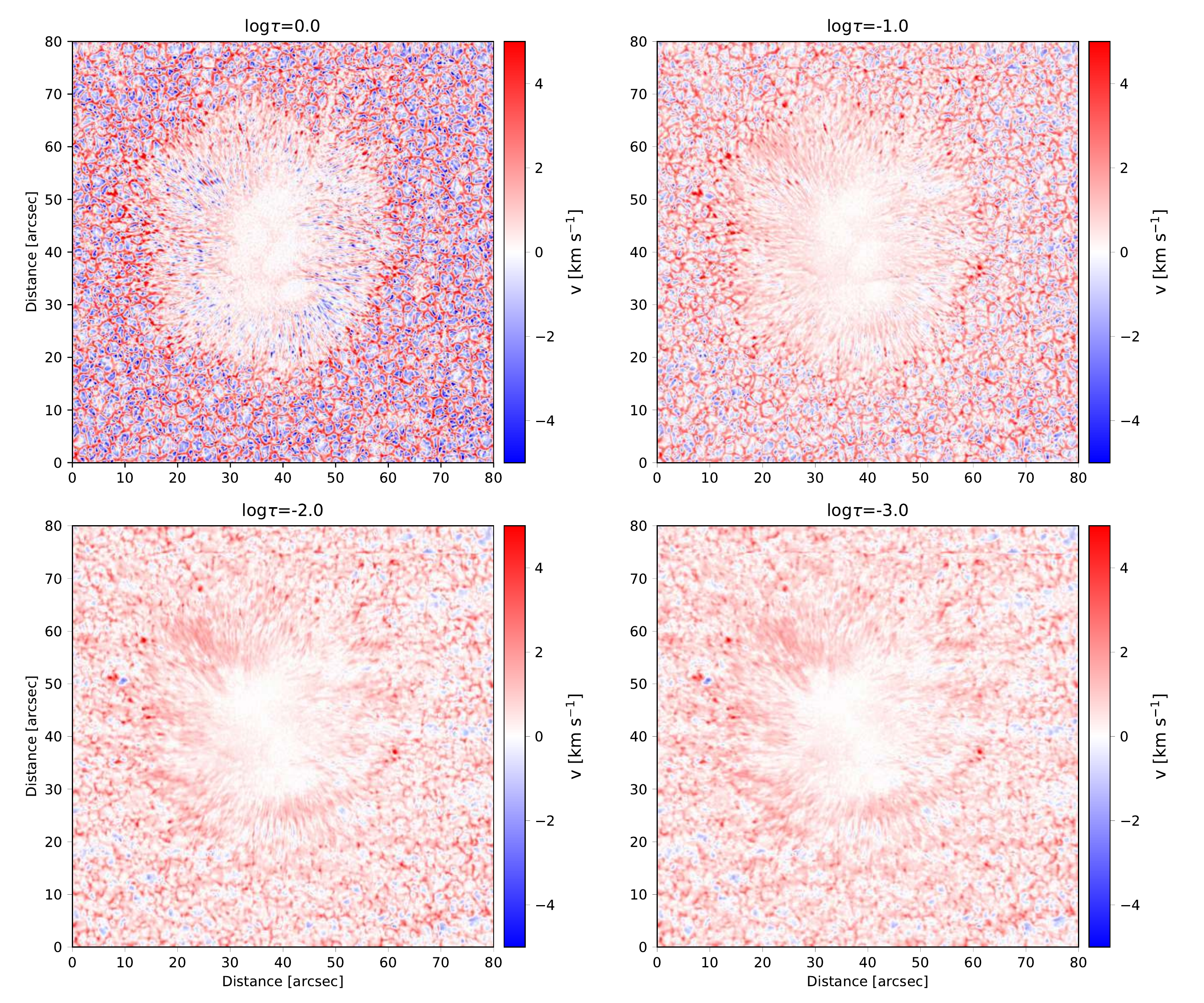}
    \includegraphics[width=0.98\columnwidth]{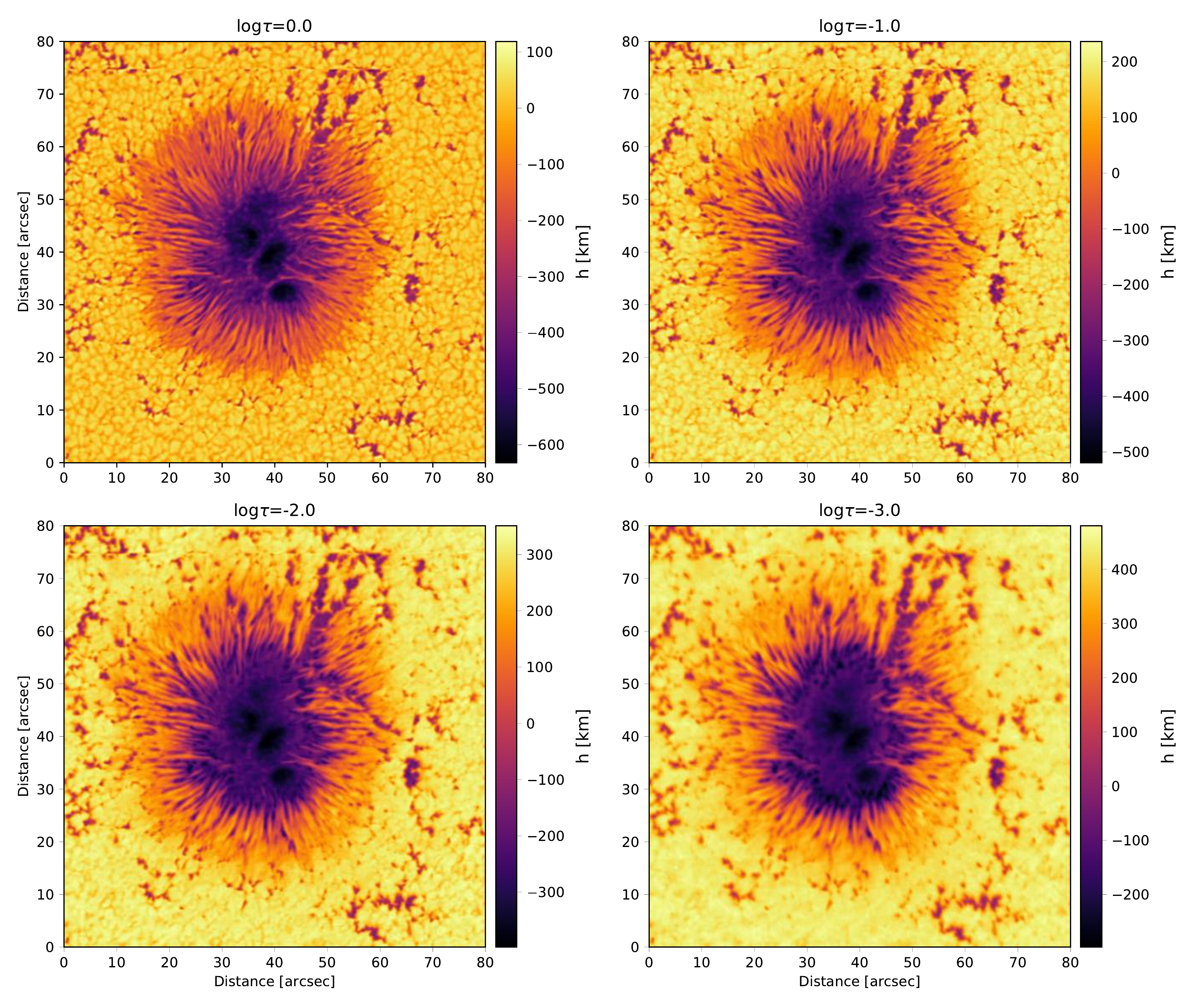}
    \includegraphics[width=0.98\columnwidth]{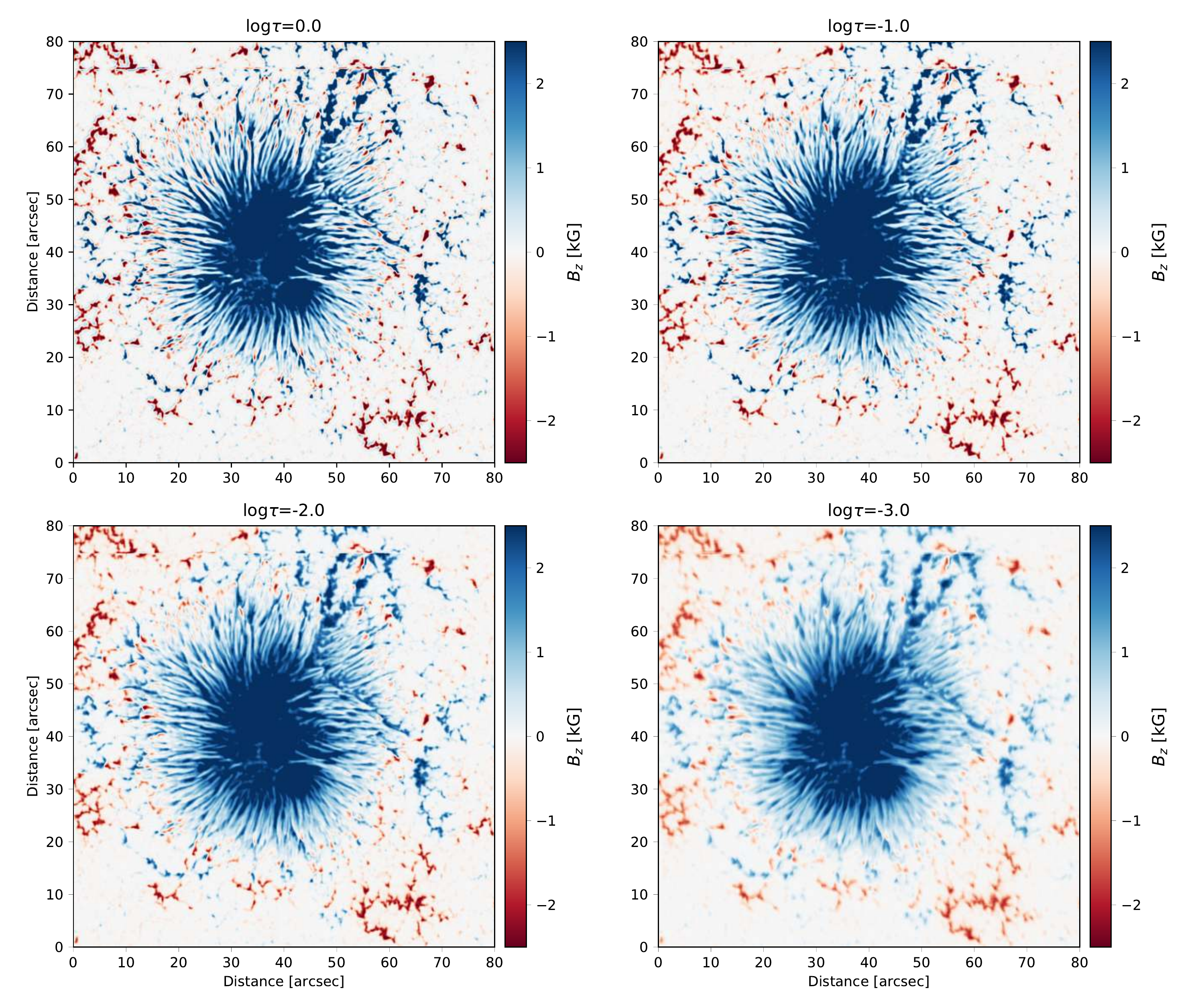}
    \includegraphics[width=0.98\columnwidth]{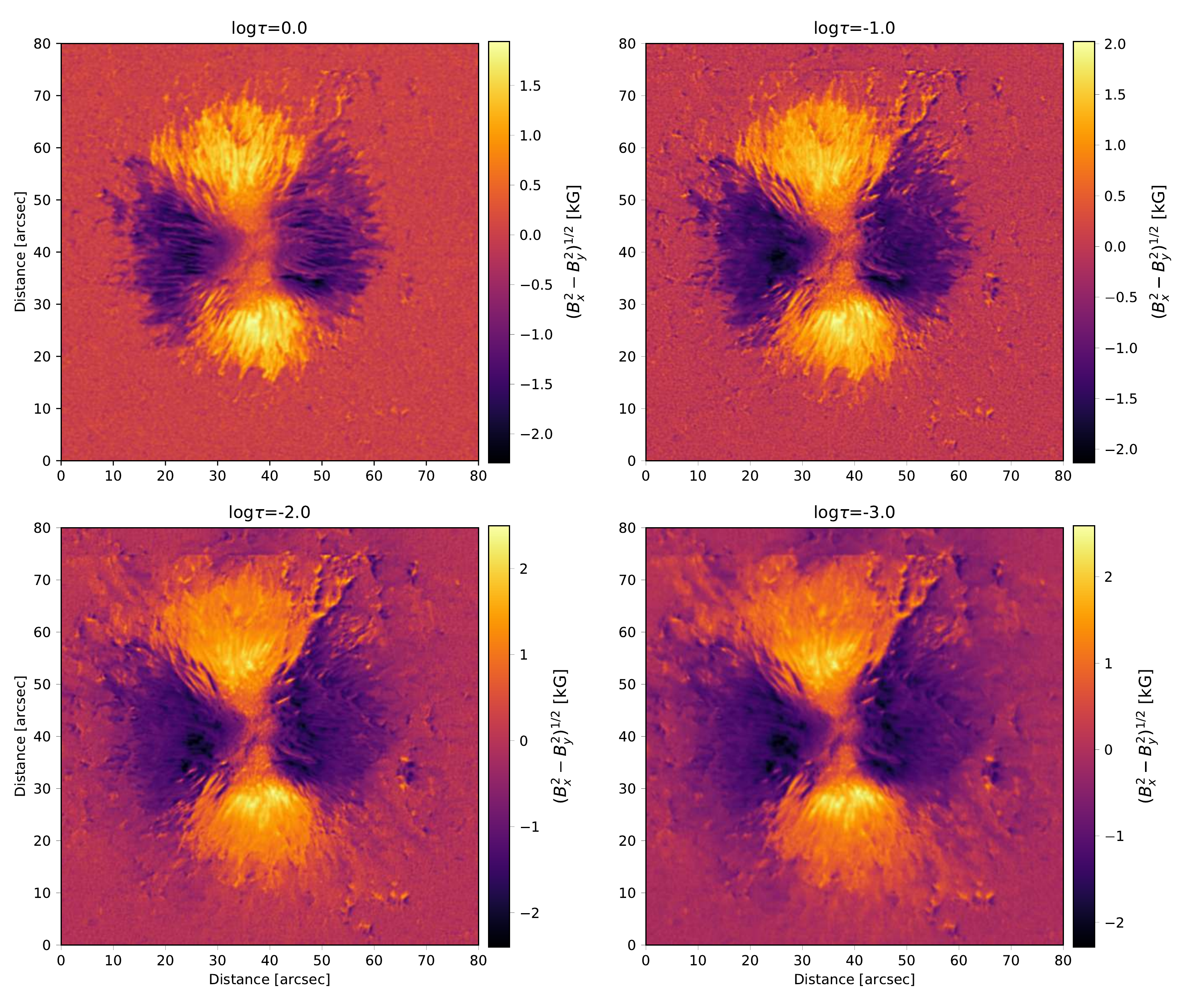}
    \includegraphics[width=0.98\columnwidth]{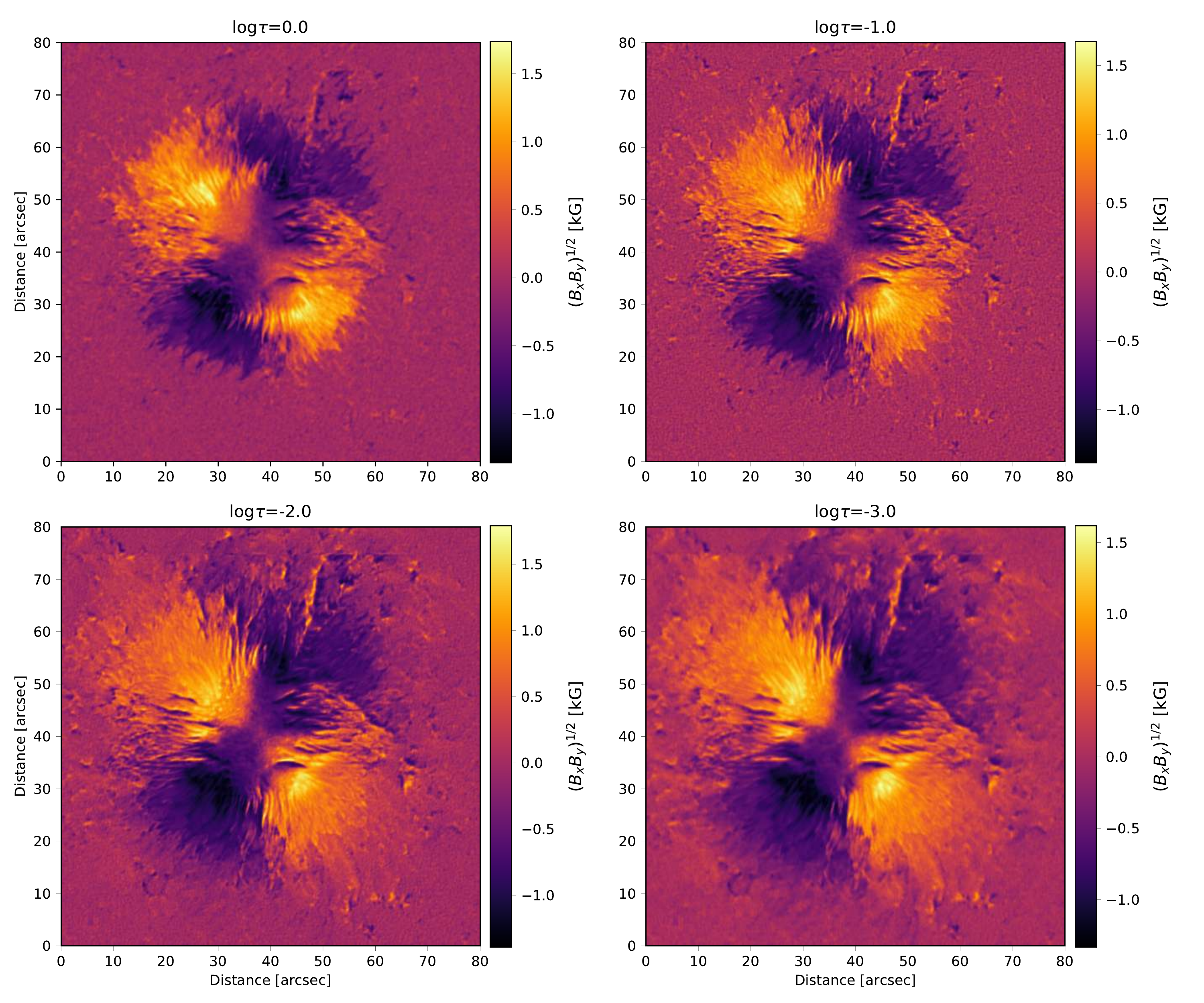}
    \caption{Predicted physical properties of AR10933 using the concatenate network.
    \label{fig:maps_ar10933b}}
\end{figure*}

%\begin{figure*}[!ht]
%    \centering
%    \includegraphics[width=2\columnwidth]{z2ejemplo2X.pdf}%
%    \caption{Caption. \comment{CARLOS: Es una red 1D con filtros 1x1. Algo que captura mejor son por ejemplo las velocidades. Mira lo que ocurre en logtau=0 con respecto a nuestras redes. Tengo la impresion que nuestras redes usan la info espacial para calcular la velocidad y no el perfil en longitud de onda.}}
%    \label{fig:maps_ar10933c}
%\end{figure*}

\begin{figure}
    \centering
    \includegraphics[width=\columnwidth]{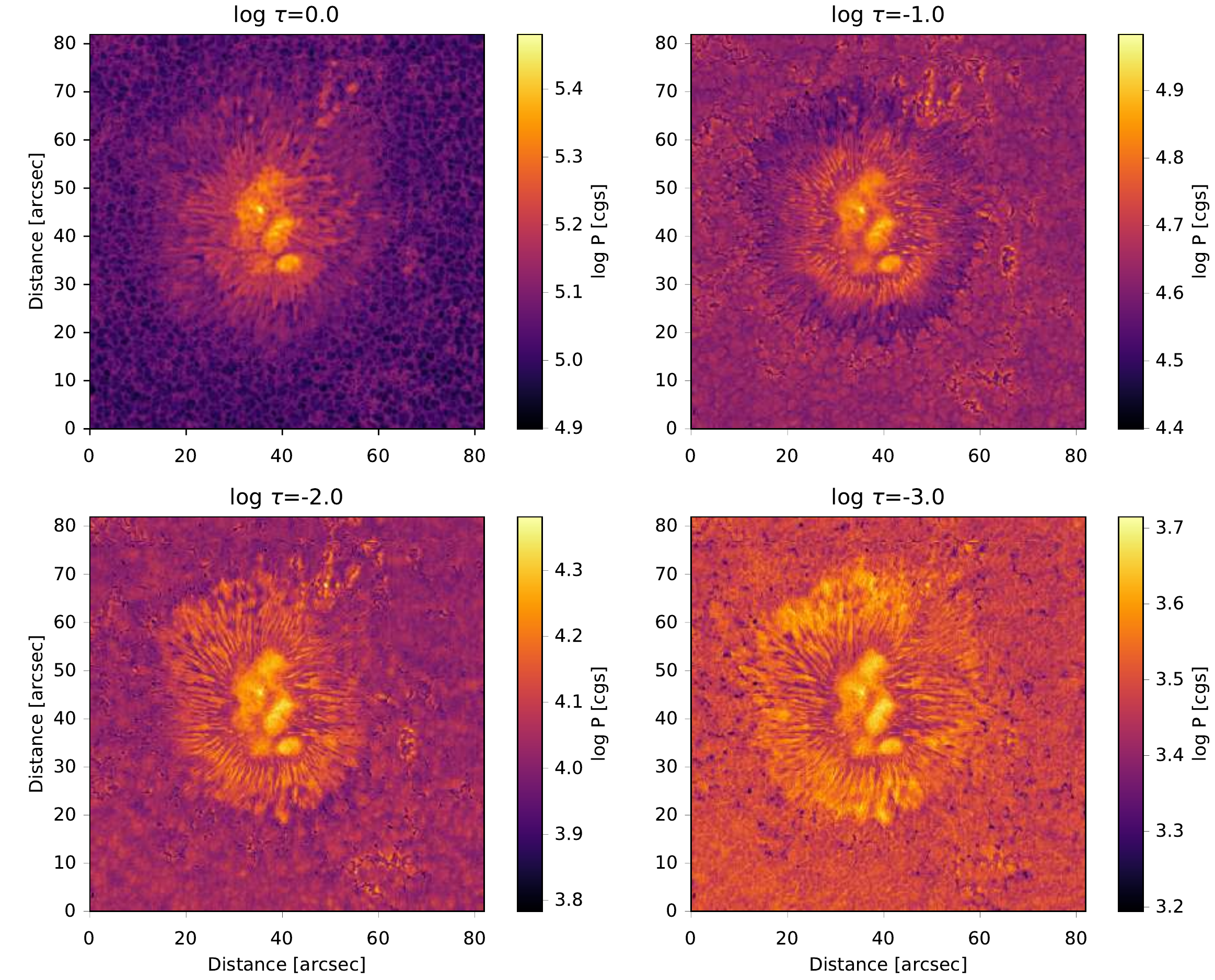}
    \caption{Total gas pressure at four different optical depth surfaces in the 
    encoder-decoder network. The pressure
    is measured in cgs units (dyn cm$^{-2}$).}
    \label{fig:pressure}
\end{figure}

\begin{figure}
    \centering
    \includegraphics[width=\columnwidth]{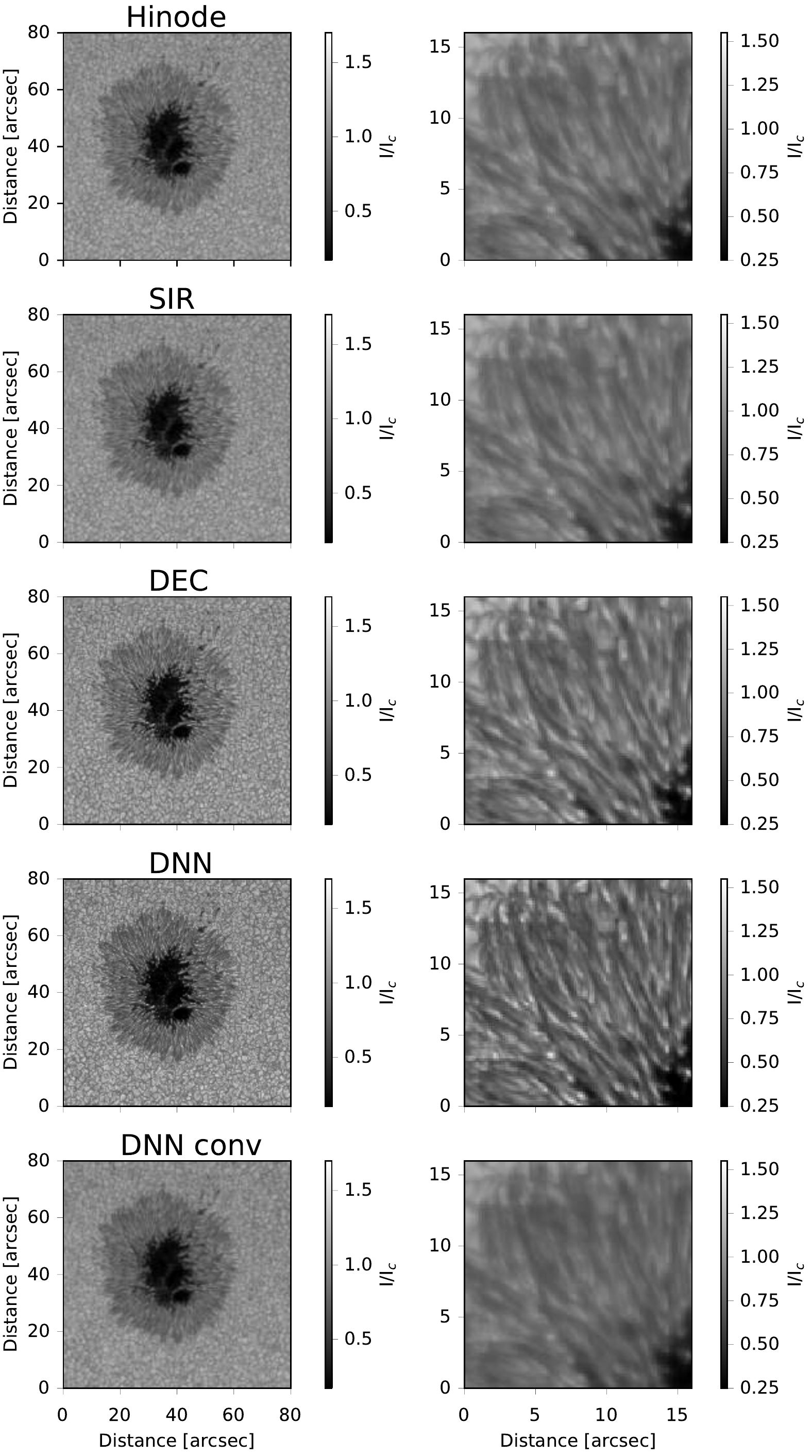}
    \caption{Continuum images (left) and zoom of the penumbra region (right). Both architectures
    provide almost identical results. The first row shows
    the original Hinode observations. The second row shows the synthetic continuum image in the
    SIR inversions. The third row refers to the data after the PCA deconvolution process.
    The fourth row is the synthesis in the model inferred by the neural network.
    The last row shows the effect of convolving with the Hinode PSF.}
    \label{fig:detail_ar10833}
\end{figure}

\subsection{Expected uncertainties}
We infer the expected uncertainties by applying the trained neural
networks to a different snapshot of the emerging flux simulation of
\cite{Cheung2010}, whose continuum image is displayed in Fig. \ref{fig:cont_validation}. 
The training and validation snapshots of \cite{Cheung2010} are separated
by 40 minutes in solar time, which is more than enough to produce sufficiently different
structures. Given the fully convolutional character of both architectures, we can
infer the physical properties on the whole field-of-view in one single run.
We compare the output of the neural network with the real value of the physical 
conditions extracted from the simulation.
Figure~\ref{fig:histograms_validation} shows the kernel
density estimate of this difference in the encoder-decoder architecture
for different optical depth surfaces. 
For comparison, we also show in Fig.~\ref{fig:histoSIR}
the kernel density estimate of the difference between the physical parameters 
of the simulation and those inferred by a standard SIR inversion of the synthetic
Stokes parameters. 
The closer to zero, the better our prediction is. The median absolute deviation 
(MAD; a robust estimation of the standard deviation) is also quoted on
each panel. We only quote that associated with $\log \tau=-1$ as representative. 
However, a detailed analysis of the width
of the distributions of Fig.~\ref{fig:histograms_validation} reveals a
weak dependence with the sensitivity of the pair of Fe \textsc{i} lines, with
slightly broader distributions in higher layers.
The resulting uncertainties are very small, in many cases smaller
than what one gets from SIR and similar to those obtained by \citep{Carroll2008}. Additionally, 
we point out that the same results are found for
both neural networks. For its novelty, we point out that we are able to 
find the height of each optical depth surface with an uncertainty below 20~km,
which is relevant if compared with that of other approaches \citep{Puschmann10,Loptien18}.
Magnetic field components are recovered with uncertainties around 100~G in absolute
units. Finally, gas pressures are recovered with uncertainties of $\sim$0.03 dex
in cgs units.

\begin{figure*}[!ht]
    \centering
    \includegraphics[width=0.98\columnwidth]{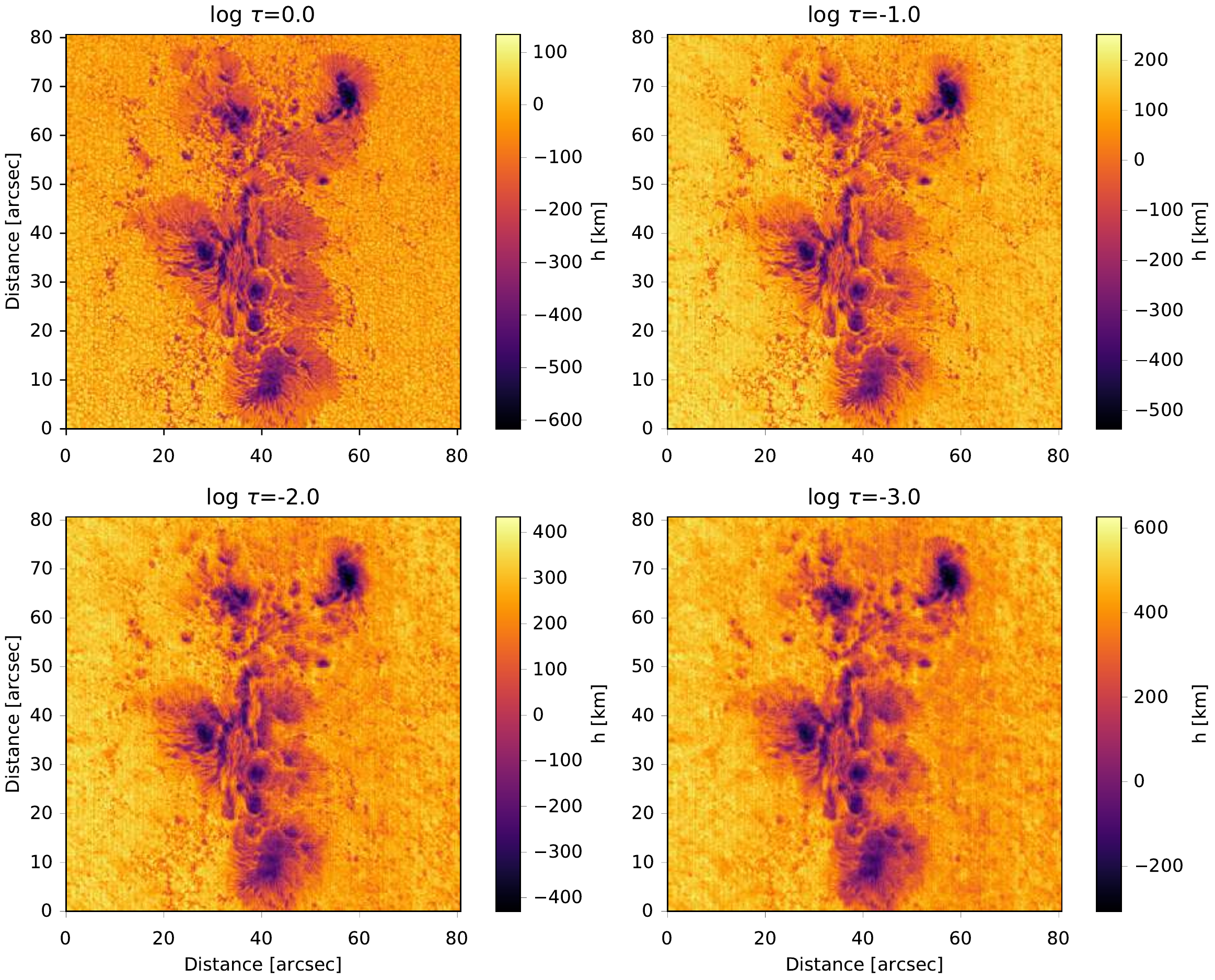}%
    \includegraphics[width=0.98\columnwidth]{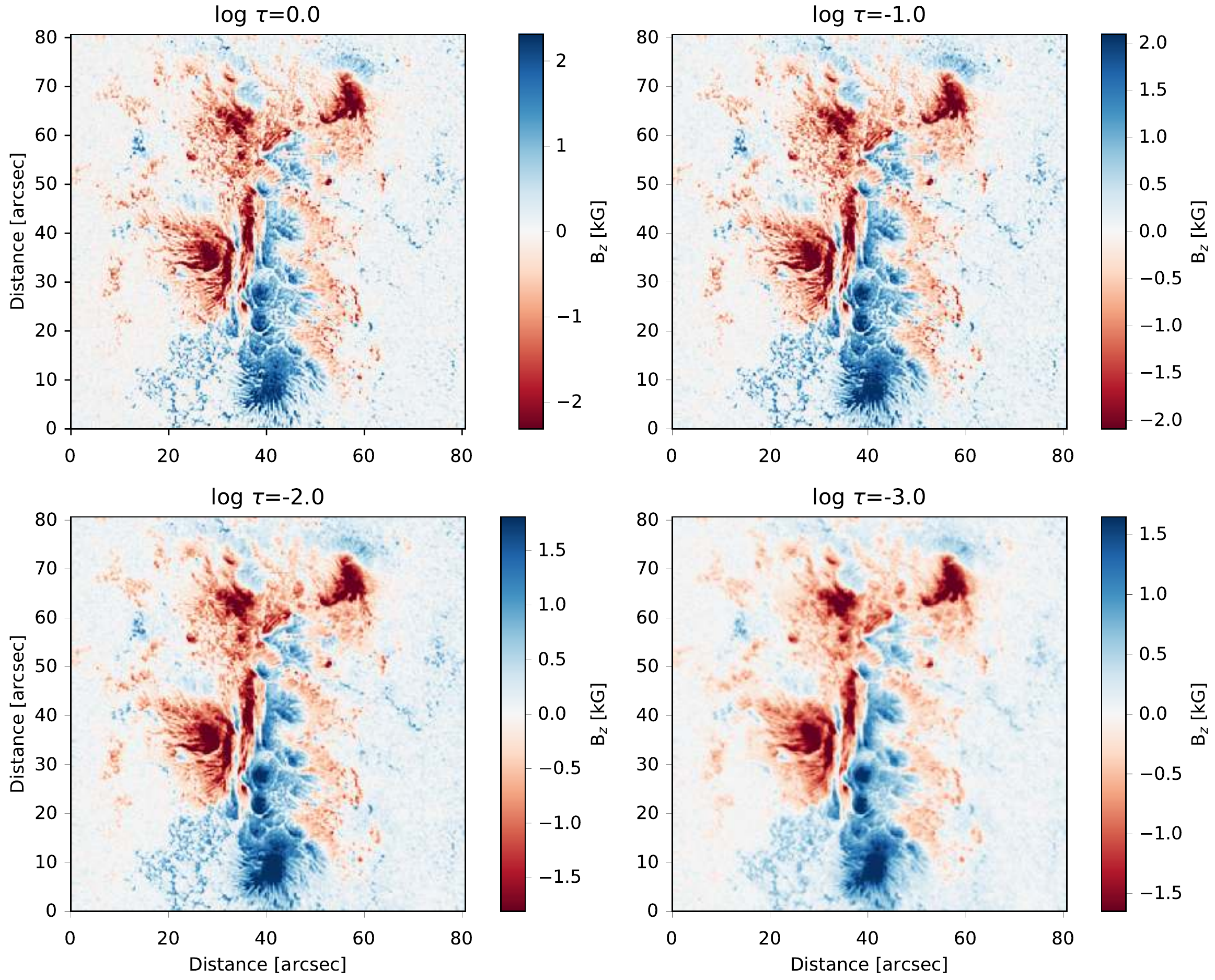}
    \includegraphics[width=0.97\columnwidth]{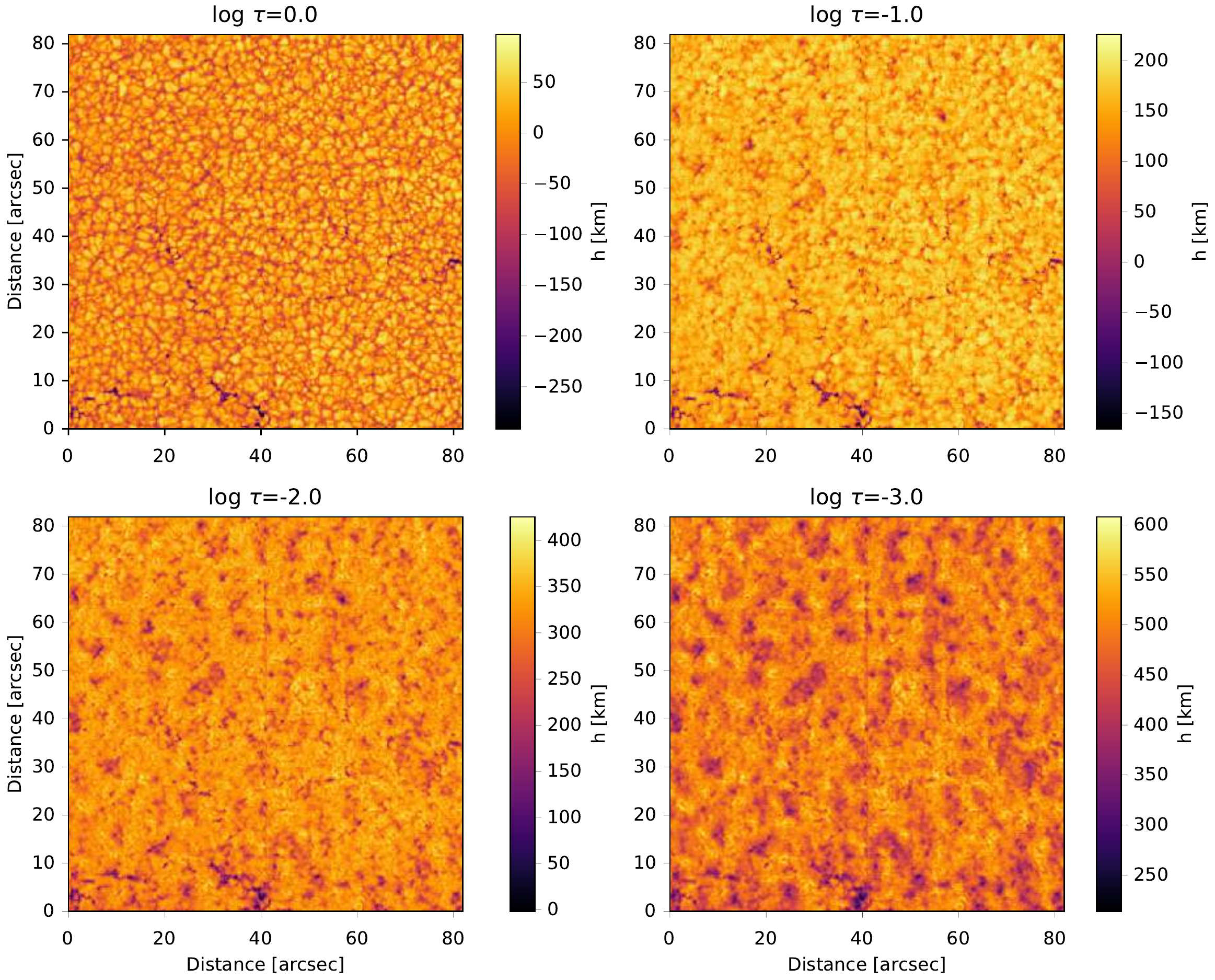}%
    \includegraphics[width=0.97\columnwidth]{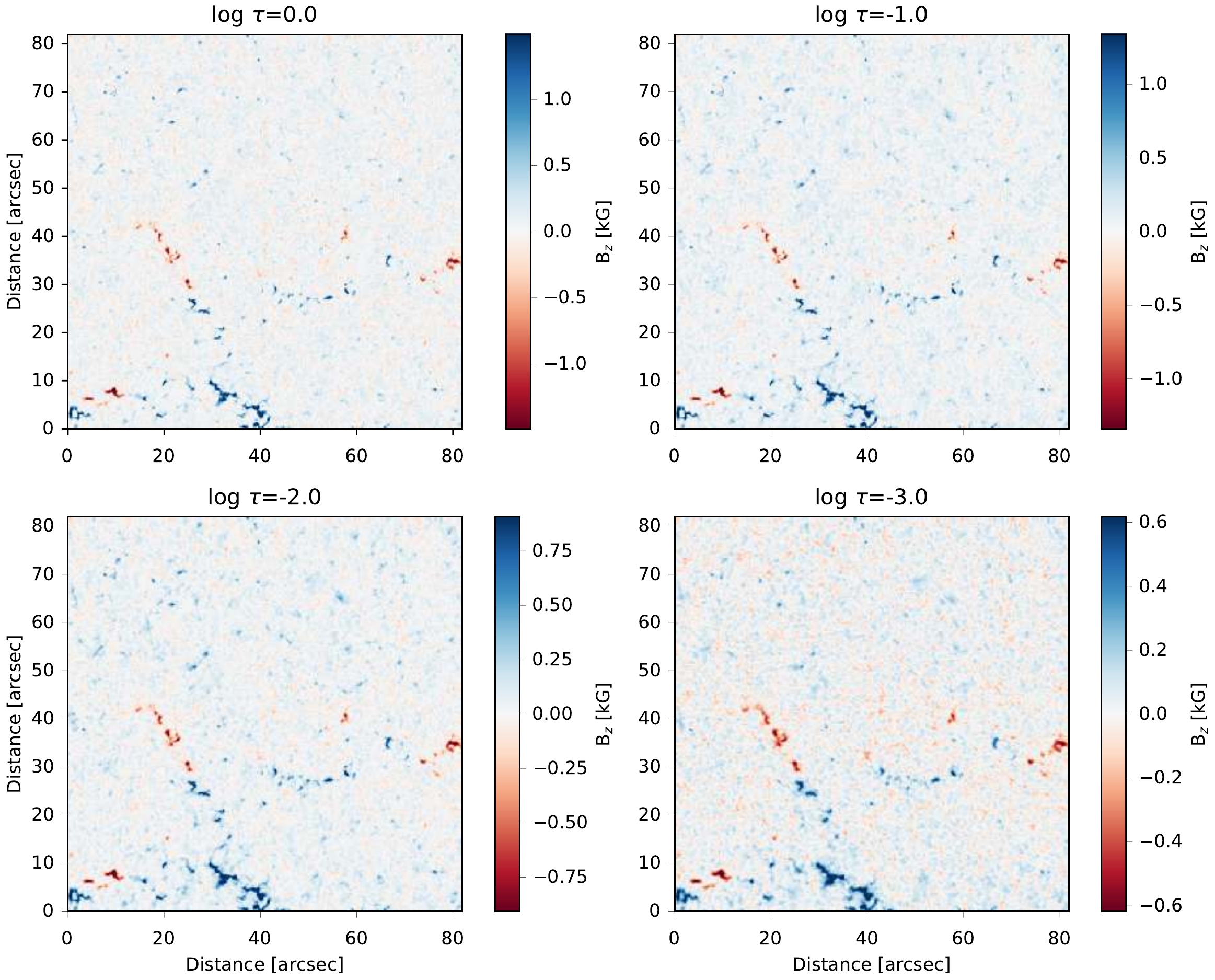}
    \caption{Some predicted properties for the complex active region 
    AR11429 and the quiet Sun with the encoder-decoder architecture.}
    \label{fig:maps_ar11429_qs}
\end{figure*}

\subsection{A simple sunspot}
We focus first on a simple sunspot, similar to that used for training.
We select the AR10933, observed with Hinode SOT/SP on 2007 January 5 from
11:20 to 14:13 UT. The sunspot is located at coordinates $(-217$\arcsec$,-13$\arcsec$)$, which 
gives an heliocentric angle of $\mu=\cos \theta=0.97$. The output of the
neural networks at four different optical depth heights 
are displayed in Fig.~\ref{fig:maps_ar10933} for
the encoder-decoder architecture and in Fig.~\ref{fig:maps_ar10933b}
for the concatenate architecture. The temperature
shows normal granulation at $\log \tau=0$, becoming more diffuse 
at the height where $\log \tau=-1$. This also happens in the simulation,
probably a consequence of the strong faculae around the sunspot.
As a consequence, the inverted granulation that we obtain with 
the neural network approach is not much contrasted.
Apart from that, bright structures (brighter than
the surrounding quiet Sun) associated with the faculae around the sunspot are seen 
at high layers. In general, the structure seen in the umbra as umbral dots
or light bridges lose contrast in higher layers.
Concerning the LOS velocity, we find the umbra practically at rest and
the granulation pattern clearly appears for heights above $\log \tau=0$.

Of special interest is the inference of the height of each pixel. We obtain
Wilson depressions of the $\log \tau=0$ surfaces in the darkest regions 
of the umbra as low as $-600$~km, similar to those quoted by \cite{Loptien18}. Light
bridges and penumbrae turn out to be much shallower, with depressions of the
order of -200~km at deep layers. In upper layers the difference turns out to
be less important, so that the height of the $\log \tau=-3$ surface is similar in the 
outer regions of the penumbra than in the surrounding quiet Sun.
The Wilson depression of the surrounding faculae is similar to that
of the penumbra, suggesting that the magnetic evacuation is similar.
On average, the height of the $\log \tau=-3$ on the quiet Sun around the sunspot
is, on average, around 430~km above that of $\log \tau=0$, something typical for the
pair of lines at 6301$-$6302~\AA\ \citep[see, e.g.,][]{Khomenko2007,Danilovic2010}.

Concerning the magnetic field information, we show $B_z$
and the two (signed) combinations of $B_x$ and $B_y$ that do not suffer
from the 180$^\circ$ ambiguity in Fig.~\ref{fig:maps_ar10933}. 
The longitudinal field shows very
concentrated magnetic field in the darkest regions of the umbra, with a 
relatively strong (still of the order of 2~kG) hazy structure in the rest
of the umbra. The maps inferred by the concatenate networks are slightly
more constrasted in $B_z$. The surrounding
faculae is of opposite polarity and a clumpy ring of opposite polarity 
surrounds the sunspot, suggesting that the magnetic field is 
returning back to the photosphere well beyond the penumbral border.
The LOS magnetic field in higher layers displays a more diffuse appearance, with
field concentrations occupying more area and with weaker fields.
The transverse components displays the typical structure with four lobes, rotated
by 45$^\circ$ from one component to the other. Appreciable signal is only seen in 
some parts of the umbra and, fundamentally, on the penumbra. Only some parts
of the faculae do show appreciable transverse components of the field. In fact,
the expansion of the magnetic field of the sunspot is much better visualized in the 
transverse component and its variation with height.

Figure~\ref{fig:pressure} shows the total gas pressure at the same four
different optical depth surface. Perhaps one of the most interesting observations
is the structure of all magnetic field concentrations away from the sunspot above
$\log \tau=0$. The gas
pressure at the central part of the magnetized region is roughly equivalent to that of the
surrounding quiet Sun at the same optical depth. However, because of the Wilson depression produced
by the rarefaction induced by the magnetic field, the radiation is coming from deeper 
layers. This behavior is different in the borders, with the Wilson depression becoming 
almost negligible and the total gas pressure strongly increasing until the quiet Sun
is reached. This whole picture can be understood as a consequence of a flux tube
with a canopy as follows. The central part of the concentration is strongly magnetized so 
the magnetic pressure is large and the gas pressure is low even if the Wilson
depression is large. On the surroundings of the flux tube the magnetic pressure
is negligible. In the interface between the two regimes there is still 
some magnetization which slightly evacuates the plasma and makes it 
possible to observe deeper layers, where the pressure is larger than in the
surrounding quiet Sun.

One of the desirable consequences of the neural networks is that the
physical quantities in the output are decontaminated from 
the PSF. This is clearly seen in Fig.~\ref{fig:detail_ar10833}, where we
show continuum images for the whole region (left column) and for a detail
in the penumbra (right column). These results are indistinguishable
in the two architectures. The root-mean-square contrast, computed as 
the ratio between the standard deviation and the mean of the brightness in 
the continuum in a quiet Sun region, is 
6.25\% for the original Hinode observations. SIR inversions provide a very
similar contrast of 6.23\%. The contrast obtained after deconvolution
goes up to 8.75 \%. This contrast can be a little higher
if more iterations of the Lucy-Richardson algorithm are applied, but we have
preferred to be in the conservative side to avoid introducing 
artifacts. The contrast obtained from the
synthesis in the models inferred from the neural network goes up to 13.58\%, very close to the contrast quoted
for high-resolution synthetic observations obtained from
3D MHD simulations \citep[e.g.,][]{Danilovic2008}. As a test for consistency, when the
continuum synthetic image is convolved again with the Hinode PSF, we obtain
a contrast of 6.09\%.

\subsubsection{A complex sunspot}
We have also analyzed a much more complex group of sunspots, 
that correspond to the AR11429. This observation was acquired with
Hinode SOT/SP on 2012 March 6 from 22:10 to 22:43 UT. The sunspot is 
located at coordinates $(-554$\arcsec$,-371$\arcsec$)$, which corresponds to an heliocentric 
angle of $\mu=\cos \theta=0.72$. The heliocentric angle is perhaps too large
in this case for the results to be quantitatively relevant given that the training set
was obtained using only synthesis at disk center. We anticipate that
a suitable training set with synthesis carried out at different
heliocentric angles could be used to quantitatively analyze observations
like this one. The interest of showing this
region is to show that the DNN approach can seamlessly be used to infer the
physical properties in very complex regions. The Wilson depression and
$B_z$ are displayed
in the upper two rows of Fig.~\ref{fig:maps_ar11429_qs} while additional figures with
the rest of physical quantities can be found in the repository created for this
paper. These Wilson depressions might differ from the real ones when measured perpendicular
to the solar surface. Concerning the LOS magnetic field, we find a very complex
region with three large-scale polarity inversion lines that are captured
by the inversion. Additionally, we find small regions of opposite polarity
embedded inside regions of negative polarity.

%\begin{figure*}
%    \centering
%    \includegraphics[width=\textwidth]{ar10933_2019-02-28-10:07_-lr_0_0003_ar10933_detail_Bx.pdf}
%    \includegraphics[width=\textwidth]{ar10933_2019-02-28-10:07_-lr_0_0003_ar10933_detail_Bz.pdf}
%    \caption{Same of Fig. \ref{fig:comparison_1} but for $B_x$ and $B_z$.}
%    \label{fig:comparison_2}
%\end{figure*}

% \begin{figure*}
%     \centering
%     \includegraphics[width=\textwidth]{z7ejemplo1_X.pdf}
%     \includegraphics[width=\textwidth]{z7ejemplo2_X.pdf}
%     \includegraphics[width=\textwidth]{z7ejemplo3_X.pdf}
%     \caption{\comment{Quiza me puedes pasar tus datos de la inversion con NN y lo pego
%     en la figura \ref{fig:comparison_1} como un panel mas.} Me parece genial la idea, asi no repetimos tantas veces lo mismo!}
%     \label{fig:comparison_2b}
% \end{figure*}

\subsubsection{Quiet Sun}
Finally, we show some results with the observation acquired with Hinode SOT/SP on 2007 March 10 from 
11:37 to 14:37 UT at disk center. This is a very large portion of the quiet Sun and we only show
a small patch of the observation. The results are shown in the lower two rows of Fig.~\ref{fig:maps_ar11429_qs}. The 
map of $B_z$ displays a large part of
the internetwork with very weak fields and a filamentary network structure. The
expansion of the field with height is clearly inferred, together with a significant reduction in 
amplitude. Concerning the Wilson depressions, the $\log \tau=0$ surface on the
network is only $\sim$150~km below that of the internetwork. In the internetwork, the
Wilson depression for $\log \tau=0$ has a value of $-4.16\pm37.3$ km, while
it becomes $152.9 \pm 30.5$~km, $314.5 \pm 34.7$~km and $486.5 \pm 37.0$~km for
$\log \tau=-1$, $\log \tau=-2$ and $\log \tau=-3$, respectively. Consequently, the
corrugation of the constant optical depths surfaces is of the order of 35~km in
very weakly magnetized regions. 

\begin{figure*}
    \centering
    \includegraphics[width=0.90\textwidth]{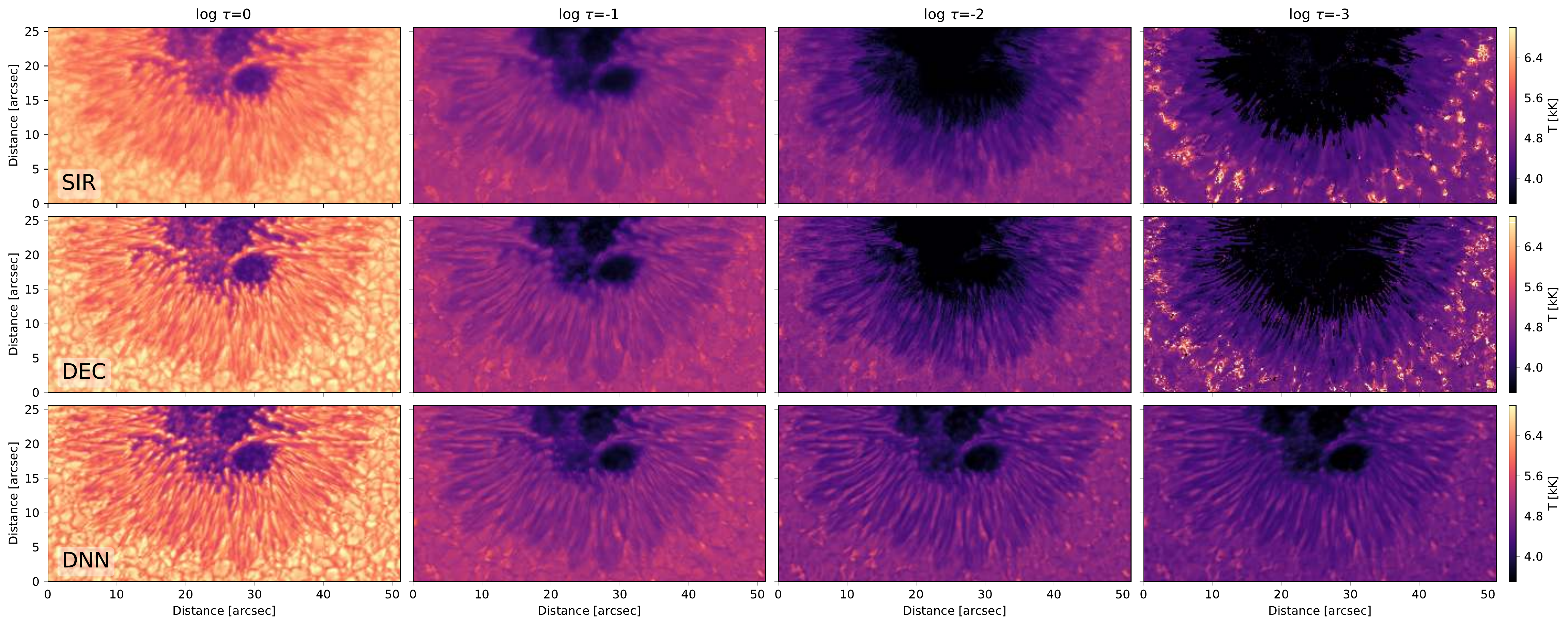}
    \includegraphics[width=0.90\textwidth]{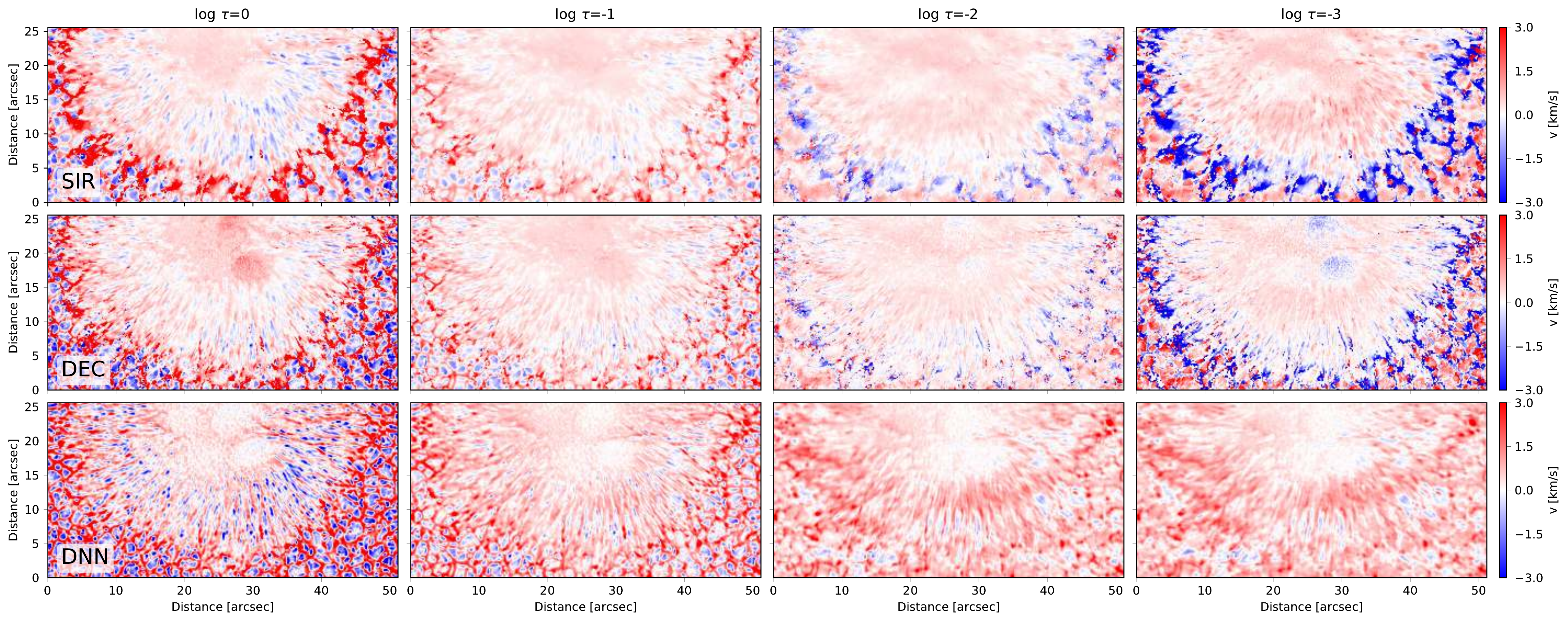}
    \includegraphics[width=0.90\textwidth]{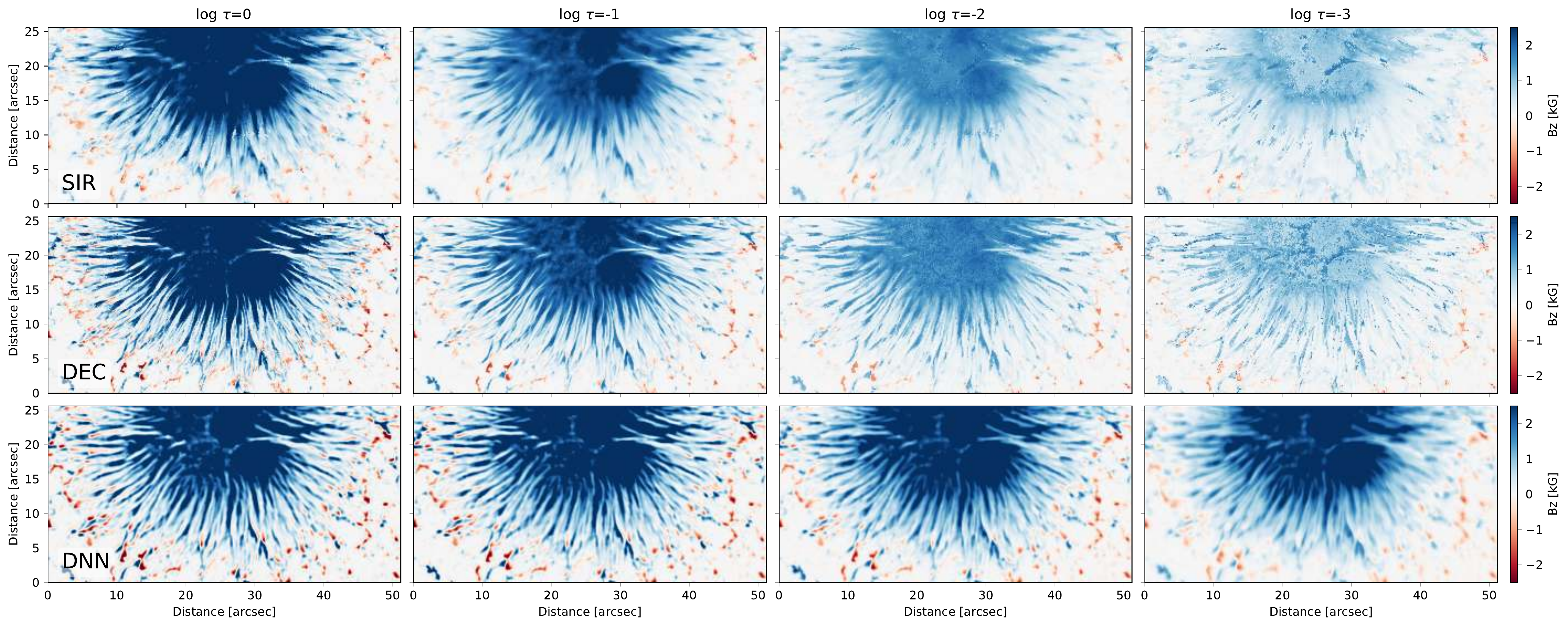}
    \caption{Detailed comparison of results from the neural network approach (in this case
    with the concatenate architecture) and a standard SIR inversion.
    We show results for temperature (top three rows), LOS velocity (middle three rows)
    and longitudinal field (bottom three rows). Inside
    each set of panels, the top row shows the results of SIR inversions at four selected optical
    depth heights, the middle row shows the results of SIR inversions after deconvolving the 
    Stokes profiles, while the bottom row shows the results of the neural approach.}
    \label{fig:comparison_1}
\end{figure*}

\subsection{Comparison with standard inversions}
It turns out difficult to compare the output of the neural network with standard inversions
because the output of both methods is fundamentally different. In the case
of standard SIR inversions, one obtains the average empirical atmosphere that 
produces a good fit to the profiles in a pixel. In the case of the neural approach, 
the training was done by injecting the Stokes profiles degraded with the PSF
and producing the physical conditions rebinned to the spatial resolution
of Hinode. The inferred model atmospheres from both approaches do not need to 
be either similar or compatible.

\begin{figure*}
    \centering
    \includegraphics[width=0.9\textwidth]{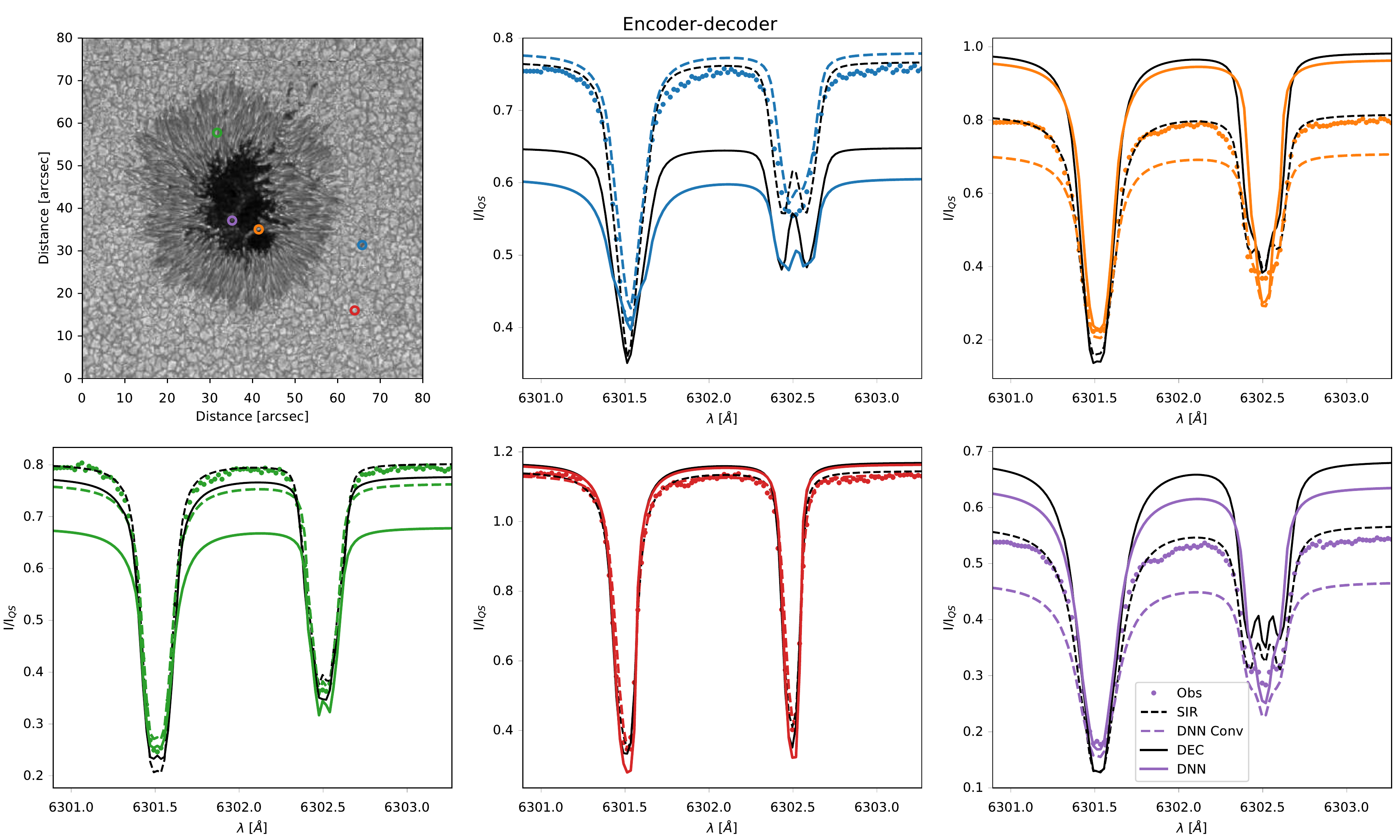}
    \vspace{0.5cm}
    \includegraphics[width=0.9\textwidth]{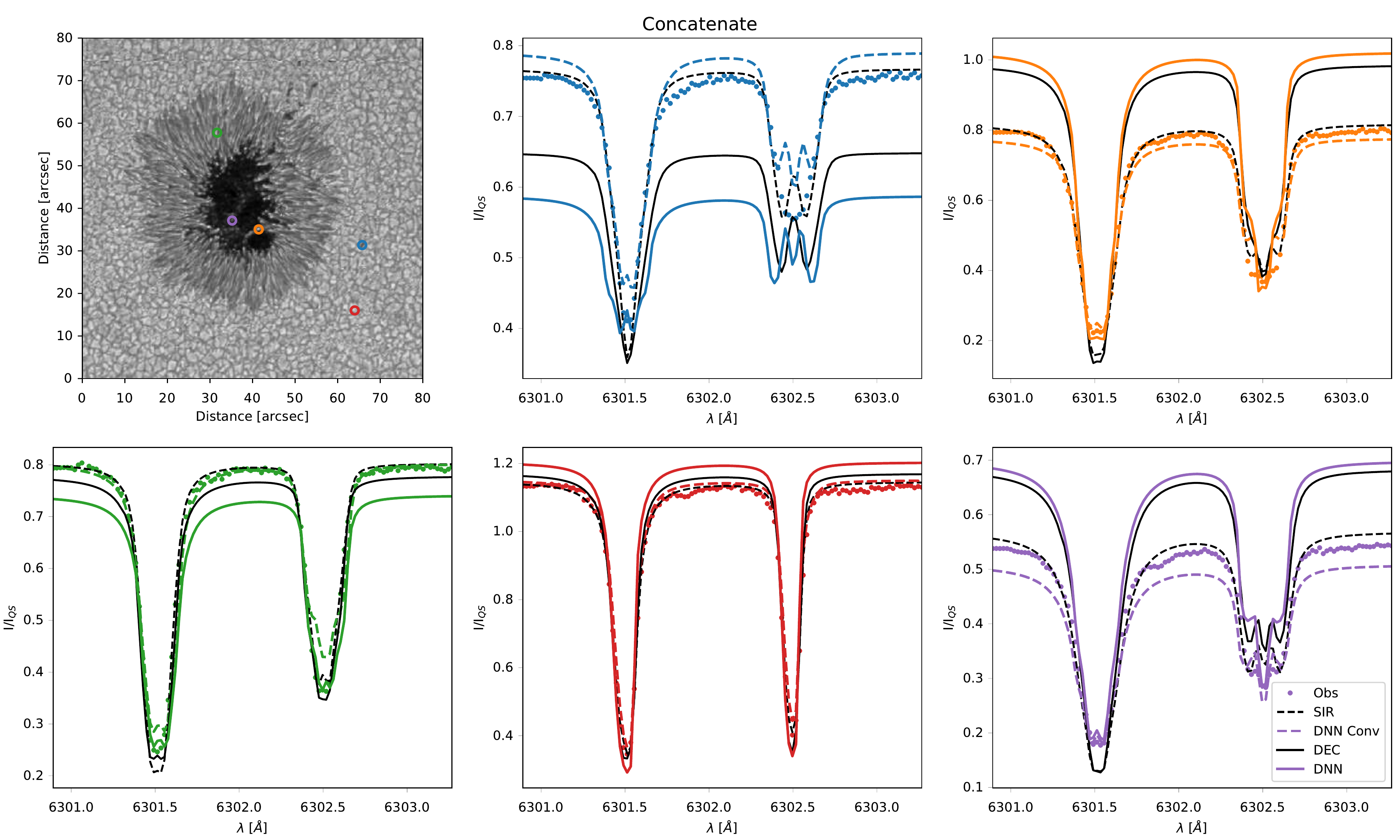}
    \caption{Comparison of Stokes $I$ profiles for all considered inversion approaches in five
    representative pixels. The upper two
    rows show the results in the encoder-decoder architecture. The lower two rows
    correspond to the concatenate architecture. The result of the SIR inversion is shown in black
    dashed line and the inversion obtained with the deconvolved data in black solid
    line. The profile synthesized in the neural approach is shown in solid
    color line (in different color for each panel) and the result convolved again
    with the Hinode PSF in dashed color line.}
    \label{fig:stokesI}
\end{figure*}

\begin{figure*}
    \centering
    \includegraphics[width=0.9\textwidth]{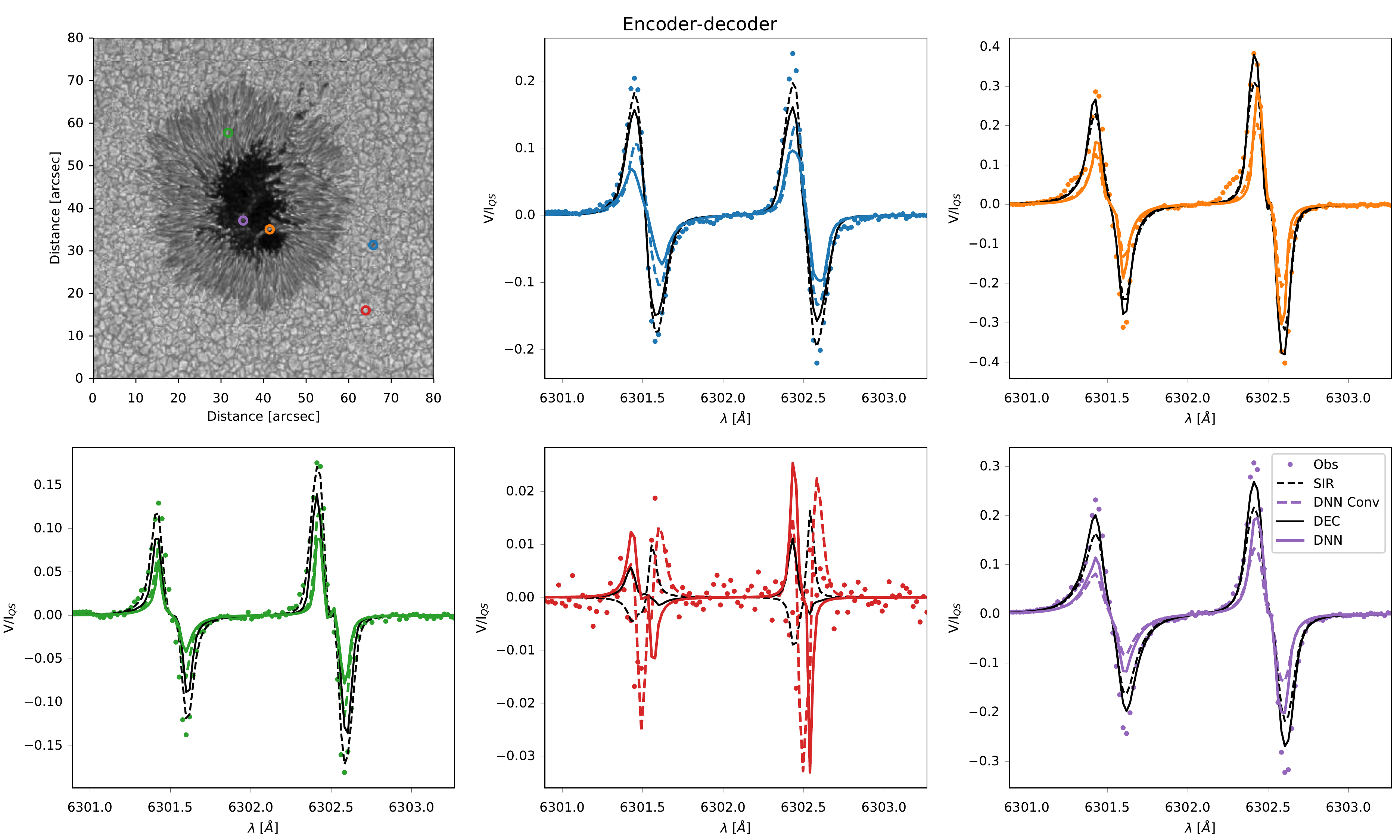}
    \vspace{0.5cm}
    \includegraphics[width=0.9\textwidth]{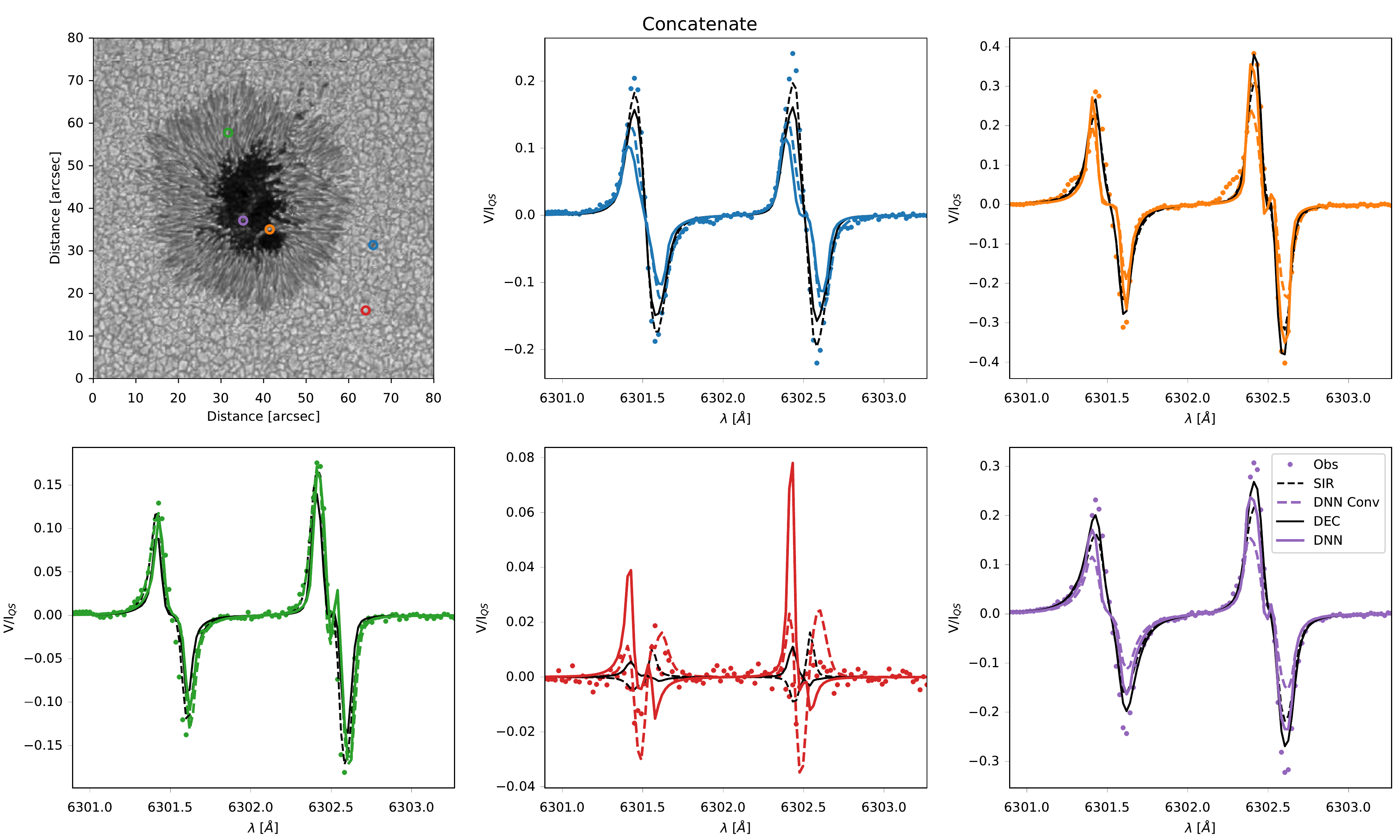}
    \caption{Like Fig.~\ref{fig:stokesI} but for Stokes $V$.}
    \label{fig:stokesV}
\end{figure*}

We apply SIR\footnote{The version used for
the inversion of these maps can be found on \url{https://github.com/cdiazbas/SIRcode}.
A paralellized implementation of SIR using Message Passing Interface (MPI) can be found 
on \url{https://github.com/cdiazbas/MPySIR}.} to carry out the inversion of the observed
Stokes profiles. We carried out a pixel-by-pixel inversion with an initialization 
of the inclination of the magnetic field given by the Stokes $V$ polarity 
($\theta_B=$45$^\circ$ if the blue lobe is positive and $\theta_B=$135$^\circ$ otherwise). 
We choose the 
{Harvard-Smithsonian Reference model Atmosphere} \citep[HSRA,][]{HSRA1971}, covering the 
optical depth range $-5.0 < \log(\tau) < 1.4$, as an initial guess atmospheric model. 
We add a magnetic field strength of 500\,G to the initial guess. The inversions were 
performed in three cycles. The temperature 
stratification $T(\tau)$ was inverted using 2,\,3, and 5 nodes in each cycle, respectively.
Two nodes were used for magnetic field strength $B(\tau)$, magnetic 
field inclination $\theta_B(\tau)$, 
and line-of-sight velocity v$(\tau)$. The magnetic field azimuth $\phi_B$ was set 
constant with height. 
Although a more complex configuration can be used, we have tried to be conservative 
as we only have two spectral lines with similar line formation regions. Only partial
information about the stratification of the magnetic field can be extracted.

Figure~\ref{fig:comparison_1} shows a detail in the penumbra for
the temperature, LOS velocity and LOS magnetic field. The results obtained with a SIR inversion in the
original Hinode profiles are displayed in the upper row, the SIR inversions 
applied to deconvolved data are displayed in the middle row. Finally, the output of
the neural approach (in this case with the concatenate architecture)
is displayed in the last row. The increased contrast that
we discussed in the continuum images of Fig.~\ref{fig:detail_ar10833} is also
present in the temperature maps. Filaments are clearly visible in 
temperature even at $\log \tau=-3$, something that also happens in standard 
inversions after deconvolution but with a much lower contrast.
The salt-and-pepper artifacts in the inversion at higher layers 
does not appear in the neural approach, which gives a perfectly smooth
map given the prior information acquired during the training.
Concerning the velocity, we very clearly detect the granulation pattern in deeper
regions, which largely disappears at $\log \tau=-3$. In general, we find
blueshifts in all the brightest elongated filaments of the penumbra, with redshifts 
on the laterals, similar to the result of \cite{Tiwari2013}. The results are not
so clear in the original data inverted with SIR but becomes more conspicuous in the 
inversions after deconvolution.
Finally, the discrepancy between the inversion 
and the neural network in the LOS magnetic field 
at higher layers is caused by a combination of the insensitivity of the lines to the field at
such heights and to the fact that we are training with a single
sunspot. However, it is also important to point out that our approach
is more powerful than a pixel-by-pixel inversion because it takes into account
the presence of correlation in the spatial and the $\log \tau$ dimensions.

\begin{figure*}
    \centering
    \includegraphics[width=0.89\textwidth]{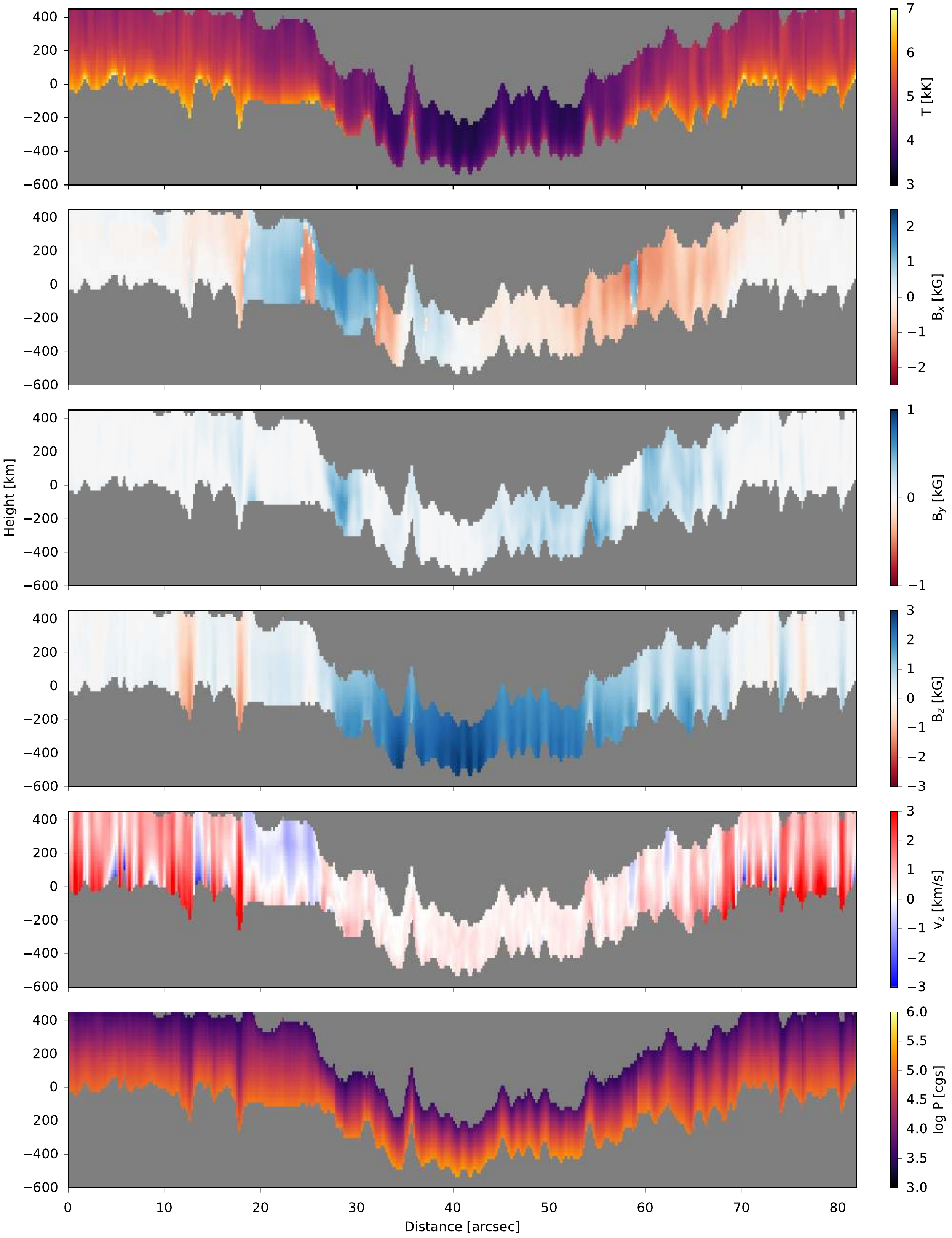}
    \caption{Physical properties displayed in geometric height for a cut at position $X=40''$
    in the map of the active region AR10933. They have been obtained with the encoder-decoder
    architecture.}
    \label{fig:light_bridge}
\end{figure*}

\subsection{Synthetic Stokes profiles}
It is important to keep in mind that, when neural networks are used for
the inversion of Stokes profiles, the resulting physical maps are not forced to
fit the profiles like in standard inversion codes. The inversion does not
proceed by defining a merit function that measures the misfit between
the synthetic Stokes profiles and the observed ones. The network learns
an approximate mapping between the Stokes profiles and the physical conditions
in an end-to-end supervised manner by minimizing the misfit between the 
inferred physical conditions and the ones used during training. We have plans
to remove this limitation in the near future, as we explain with more detail
in Sect. \ref{sec:conclusions}.

For the moment, we use SIR to synthesize the Stokes profiles emerging from the 
atmosphere inferred by the neural network and compare it with the original Stokes profiles.
To this end we use the inferred maps at the seven optical depth surfaces
and generate a smooth model using spline interpolation. This model is then
fed to SIR to return the emerging Stokes profiles. We warn the reader that this comparison is
somehow misguided and one should not conclude that our results are wrong
just because the SIR synthesis in the inferred models does not
perfectly fit the observed profiles. This behavior is due to, at least, two
reasons. The first one is that the inferred atmospheric model is compensated for
the Hinode PSF. This effect is also present in the inversion
of the deconvolved data. The second reason is that the neural
network is trained to return average atmospheric models per pixel, and we know that the synthetic
Stokes profiles in the average model is different from the average of the synthetic
Stokes profiles.

With all these caveats in mind, some examples are
displayed in Fig.~\ref{fig:stokesI} for Stokes $I$ and in Fig.~\ref{fig:stokesV} for Stokes $V$
for five representative pixels and the two architectures. We show the
original observation in colored circles, the result of the SIR inversion in black
dashed line and the inversion obtained with the deconvolved data in black solid
line. The profile synthesized in the neural approach is shown in solid
color line (in different color for each panel) and the result convolved again
with the Hinode PSF in dashed color line. This figure shows that SIR, by construction, correctly
fits the observed profiles. The profiles of the inversion of the deconvolved
maps show a much larger contrast (dark regions become darker and bright 
regions become brigther) as a consequence of the decontamination from the 
Hinode PSF. The result of the neural networks is very close to this
deconvolved result, although it often shows an even larger increase in the contrast.
For instance, the location marked in green in the penumbra shows a lower continuum
than the observed profile. When the profiles are degraded
with the Hinode PSF, we recover a profile that is close to the observation,
and also to the standard SIR inversion. 

\section{Light bridge}
Our ability to recover the inversions in geometric scale allows us to study in
detail interesting structures like the light bridge located at coordinates (40\arcsec, 35\arcsec).
To this end, we show a vertical cut at $X=40$\arcsec\ for some
relevant physical quantities in Fig. \ref{fig:light_bridge}. To generate
this plot we interpolate the information provided by the
inversion in a common geometric height axis ranging from $-600$~km to 450~km.
The vertical cut crosses the light bridge at position $\sim 35$\arcsec, roughly
in the position of the orange circle in Fig.~\ref{fig:stokesI}. The Wilson depression
for the umbra is larger than $-400$~km on average. On the light bridge the $\log \tau=0$
surface lies at a height of $-200$~km on average and the line formation
region spans roughly 250~km in height, so that the $\log \tau=-3$ surface 
is located at $\sim$50~km. The temperature stratification of the light
bridge is closer in structure to that of the quiet Sun or the penumbra, with high temperatures in the lower
boundary of $\sim$7000~K. However, slightly lower temperatures are found
in the upper layers of the light bridge as compared with those of the
quiet Sun. The cusp of the light bridge is found to be slightly
redshifted, with very extremely weak blueshifts around, suggesting a weak convection-like 
motion. Concerning the magnetic field, we find a reduced $B_z$ in the cusp, with
relatively strong horizontal components of the field in the surroundings of the light bridge.
The inferred pressure displays a very good pressure balance in general. Among the exceptions
we find the light bridge. It is in a strong pressure imbalance with the surroundings, with 
differences larger than one order of magnitude. We also find other locations
with large imbalances, but are normally associated with strong plasma
downflows in intergranular lanes.
In general, the behavior of the light bridge is very similar to that found in any
penumbra filament, clearly distinct from the surrounding umbra. 
The global picture of the magnetic structure we obtain is similar to that
inferred by \cite{Felipe2016}. 

\section{Conclusions and future work}
\label{sec:conclusions}
We have introduced and analyzed an approach for the very fast inference of physical
properties from the observation of Stokes profiles leveraging deep learning tools. 
The case we have discussed is 
specialized for Hinode observations in the pair of Fe \textsc{i} lines
at 630 nm under the assumption of LTE, but we expect applications to other cases to flourish after this work.
To this end, we release both the training and evaluation codes as open source so that they 
can be easily adapted to other applications.
The approach has a number of advantages. Firstly, it is extremely fast, especially when used with GPUs. 
It can invert maps of 512$\times$512 pixels in less than 200 ms using an off-the-shelf GPU.
Second, it can easily provide quantities that are very difficult to obtain with normal inversions,
like the Wilson depression or the gas pressure. A validation analysis 
suggests that the neural networks are able to return the physical quantities in a
3D cube with enough precision: temperatures are returned with median deviations below
$\pm$82~K, velocities with deviations below $\pm$0.41~km s$^{-1}$, heights of the 
optical depth surfaces with deviations below $\pm$18~km and gas pressure in logarithmic
units with deviations around $\pm$0.02~dex. Concerning the magnetic field, we quote deviations
that are of the order of or below $\pm$100~G. We point out that
the encoder-decoder architecture is able to better exploit the spatial
coherence of the physical variables and is slightly less affected by the 
observational noise. However, although both architectures give
good results, in general we find slightly more robust results with the 
concatenate architecture when applied to Hinode data.

Our approach in this contribution can be understood as a 
maximum a-posteriori solution to the inversion problem (data-driven priors are
introduced by the neural network during training) in which uncertainties
are not computed. 
Obviously, the final aim of any inversion code would be to fully characterize the posterior distribution
function for all observed pixels. We anticipate that this will be possible in the near future
with the use of machine learning because of the enormous interest in the community
of developing interpretable algorithms. Currently, one could argue
that the combination of CNNs with their spatial regularization capabilities
with recent developments like INNs could be a possible avenue of research. The 
posterior distributions would surely be multimodal \citep[e.g.,][]{diazbaso19b,diazbaso19c} due 
to some well-known ambiguities. However, posterior distributions inside each mode
tend to be quite well-behaved, nonetheless with the existence of 
degeneracies \citep{2006A&A...456.1159M,asensio07,asensio09,diazbaso19a}.

% \textbf{We have trained two different architectures....}

Despite the encouraging results, a caveat is in order. The fact that
the neural networks are not explicitly fitting Stokes profiles might be seen problematic.
The deep learning approach that we have used here should be considered the first
approach to the inversion of Stokes profiles. It is, though, not physically motivated
and we are convinced that introducing the radiative transfer (and perhaps
also the MHD) physics during training will
lead to significant improvements. This requires the introduction of a differentiable
forward synthesis module during training. Although we have plans to implement this idea
in the near future, we anticipate that the training will become much more time consuming.

Anyway, even if one does not trust the results of the neural networks for doing
research, it can always be used as a very high quality initial condition for 
the application of standard inversion codes. The noise filtering capabilities
of the neural network and the exploitation of spatial correlation will allow standard inversion
codes to quickly converge to very clean maps.
Additionally, the fact that our approach provides the results in geometric height opens up
the possibility of using these results with next-generation 3D Stokes inversion 
including magnetohydrostatic constraints (Pastor Yabar et al., private communication).

\begin{acknowledgements}
We are grateful to M. Rempel and M. Cheung for providing the snapshots used for
training the neural networks.
We would like to thank Nikola Vitas and Morten Franz for the help
with some synthesis of the Stokes profiles in the training sets
at the initial steps of the work. Financial support by the Spanish Ministry of Economy and Competitiveness
through project AYA2014-60476-P is gratefully acknowledged.
This project has received funding from the European Research Council (ERC) under
the European Union's Horizon 
2020 research and innovation program (SUNMAG, grant agreement 759548). The
Institute for Solar Physics 
is supported by a grant for research infrastructures of national importance from
the Swedish Research 
Council (registration number 2017-00625).
We also thank the NVIDIA Corporation for the donation of the Titan X GPU, one
of the GPUs used in this research.
This research has made use of NASA's Astrophysics Data System 
Bibliographic Services.
We acknowledge the community effort devoted to the development of the following 
open-source packages that were used in this work: 
\texttt{numpy} (\texttt{numpy.org}), \texttt{matplotlib} (\texttt{matplotlib.org}),
\texttt{Keras} (\texttt{keras.io}), and \texttt{PyTorch} (\texttt{pytorch.org}).
\end{acknowledgements}

%\begin{appendix}

%\section{Topology of the residual--block}
%We have changed the topology
%of the residual--block adding a ReLU layer after addition or before (as the 
%original study) and the result is the same.

%\end{appendix}

%-------------------------------------------------------------------

% \bibliographystyle{aa}
%\bibliographystyle{mnras}
% \b

% ibliography{general}


\begin{thebibliography}{58}
    \expandafter\ifx\csname natexlab\endcsname\relax\def\natexlab#1{#1}\fi
    
    \bibitem[{{Ardizzone} {et~al.}(2018){Ardizzone}, {Kruse}, {Wirkert}, {Rahner},
      {Pellegrini}, {Klessen}, {Maier-Hein}, {Rother}, \&
      {K{\"o}the}}]{Ardizzone2018}
    {Ardizzone}, L., {Kruse}, J., {Wirkert}, S., {et~al.} 2018, arXiv e-prints,
      arXiv:1808.04730
    
    \bibitem[{{Asensio Ramos}(2009)}]{asensio09}
    {Asensio Ramos}, A. 2009, \apj, 701, 1032
    
    \bibitem[{{Asensio Ramos} \& {de la Cruz Rodr{\'{\i}}guez}(2015)}]{Asensio2015}
    {Asensio Ramos}, A. \& {de la Cruz Rodr{\'{\i}}guez}, J. 2015, \aap, 577, A140
    
    \bibitem[{{Asensio Ramos} {et~al.}(2018){Asensio Ramos}, {de la Cruz
      Rodr{\'\i}guez}, \& {Pastor Yabar}}]{DeepMFBD18}
    {Asensio Ramos}, A., {de la Cruz Rodr{\'\i}guez}, J., \& {Pastor Yabar}, A.
      2018, \aap, 620, A73
    
    \bibitem[{{Asensio Ramos} {et~al.}(2007{\natexlab{a}}){Asensio Ramos},
      {Mart{\'\i}nez Gonz{\'a}lez}, \& {Rubi{\~n}o-Mart{\'\i}n}}]{asensio07}
    {Asensio Ramos}, A., {Mart{\'\i}nez Gonz{\'a}lez}, M.~J., \&
      {Rubi{\~n}o-Mart{\'\i}n}, J.~A. 2007{\natexlab{a}}, \aap, 476, 959
    
    \bibitem[{{Asensio Ramos} {et~al.}(2017){Asensio Ramos}, {Requerey}, \&
      {Vitas}}]{Asensio2017}
    {Asensio Ramos}, A., {Requerey}, I.~S., \& {Vitas}, N. 2017, \aap, 604, A11
    
    \bibitem[{{Asensio Ramos} {et~al.}(2007{\natexlab{b}}){Asensio Ramos},
      {Socas-Navarro}, {L{\'o}pez Ariste}, \& {Mart{\'\i}nez
      Gonz{\'a}lez}}]{2007ApJ...660.1690A}
    {Asensio Ramos}, A., {Socas-Navarro}, H., {L{\'o}pez Ariste}, A., \&
      {Mart{\'\i}nez Gonz{\'a}lez}, M.~J. 2007{\natexlab{b}}, \apj, 660, 1690
    
    \bibitem[{{Asensio Ramos} {et~al.}(2008){Asensio Ramos}, {Trujillo Bueno}, \&
      {Landi Degl'Innocenti}}]{Asensio2008}
    {Asensio Ramos}, A., {Trujillo Bueno}, J., \& {Landi Degl'Innocenti}, E. 2008,
      \apj, 683, 542
    
    \bibitem[{{Auer} {et~al.}(1977){Auer}, {Heasley}, \& {House}}]{Auer1977}
    {Auer}, L.~H., {Heasley}, J.~N., \& {House}, L.~L. 1977, \solphys, 55, 47
    
    \bibitem[{{Borrero} {et~al.}(2011){Borrero}, {Tomczyk}, {Kubo},
      {Socas-Navarro}, {Schou}, {Couvidat}, \& {Bogart}}]{Borrero2011}
    {Borrero}, J.~M., {Tomczyk}, S., {Kubo}, M., {et~al.} 2011, \solphys, 273, 267
    
    \bibitem[{{Cand{\`e}s} {et~al.}(2006){Cand{\`e}s}, {Romberg}, \&
      {Tao}}]{candes06}
    {Cand{\`e}s}, E., {Romberg}, J., \& {Tao}, T. 2006, Comm. Pure Appl. Math., 59,
      1207
    
    \bibitem[{{Carroll} \& {Kopf}(2008)}]{Carroll2008}
    {Carroll}, T.~A. \& {Kopf}, M. 2008, \aap, 481, L37
    
    \bibitem[{{Carroll} \& {Staude}(2001)}]{Carroll2001}
    {Carroll}, T.~A. \& {Staude}, J. 2001, \aap, 378, 316
    
    \bibitem[{{Cheung} {et~al.}(2018){Cheung}, {Rempel}, {Chintzoglou}, {Chen},
      {Testa}, {Mart{\'\i}nez-Sykora}, {Sainz Dalda}, {DeRosa}, {Malanushenko},
      {Hansteen}, {De Pontieu}, {Carlsson}, {Gudiksen}, \&
      {McIntosh}}]{Cheung2018_flare}
    {Cheung}, M.~C.~M., {Rempel}, M., {Chintzoglou}, G., {et~al.} 2018, Nature
      Astronomy, 173
    
    \bibitem[{{Cheung} {et~al.}(2010{\natexlab{a}}){Cheung}, {Rempel}, {Title}, \&
      {Sch{\"u}ssler}}]{Cheung10}
    {Cheung}, M.~C.~M., {Rempel}, M., {Title}, A.~M., \& {Sch{\"u}ssler}, M.
      2010{\natexlab{a}}, \apj, 720, 233
    
    \bibitem[{{Cheung} {et~al.}(2010{\natexlab{b}}){Cheung}, {Rempel}, {Title}, \&
      {Sch{\"u}ssler}}]{Cheung2010}
    {Cheung}, M.~C.~M., {Rempel}, M., {Title}, A.~M., \& {Sch{\"u}ssler}, M.
      2010{\natexlab{b}}, \apj, 720, 233
    
    \bibitem[{{Clevert} {et~al.}(2015){Clevert}, {Unterthiner}, \&
      {Hochreiter}}]{ELU2015}
    {Clevert}, D.-A., {Unterthiner}, T., \& {Hochreiter}, S. 2015, arXiv e-prints,
      arXiv:1511.07289
    
    \bibitem[{{Collados} {et~al.}(2013){Collados}, {Bettonvil}, {Cavaller},
      {Ermolli}, {Gelly}, {P{\'e}rez}, {Socas-Navarro}, {Soltau}, {Volkmer}, \&
      {EST Team}}]{EST2013}
    {Collados}, M., {Bettonvil}, F., {Cavaller}, L., {et~al.} 2013, Memorie della
      Societa Astronomica Italiana, 84, 379
    
    \bibitem[{{Danilovic} {et~al.}(2008){Danilovic}, {Gandorfer}, {Lagg},
      {Sch{\"u}ssler}, {Solanki}, {V{\"o}gler}, {Katsukawa}, \&
      {Tsuneta}}]{Danilovic2008}
    {Danilovic}, S., {Gandorfer}, A., {Lagg}, A., {et~al.} 2008, \aap, 484, L17
    
    \bibitem[{{Danilovic} {et~al.}(2010){Danilovic}, {Sch{\"u}ssler}, \&
      {Solanki}}]{Danilovic2010}
    {Danilovic}, S., {Sch{\"u}ssler}, M., \& {Solanki}, S.~K. 2010, \aap, 513, A1
    
    \bibitem[{{de la Cruz Rodr{\'\i}guez} {et~al.}(2018){de la Cruz
      Rodr{\'\i}guez}, {Leenaarts}, {Danilovic}, \& {Uitenbroek}}]{Jaime2018_Stic}
    {de la Cruz Rodr{\'\i}guez}, J., {Leenaarts}, J., {Danilovic}, S., \&
      {Uitenbroek}, H. 2018, arXiv e-prints, arXiv:1810.08441
    
    \bibitem[{{D{\'\i}az Baso} \& {Asensio Ramos}(2018)}]{Enhance18}
    {D{\'\i}az Baso}, C.~J. \& {Asensio Ramos}, A. 2018, \aap, 614, A5
    
    \bibitem[{{D{\'\i}az Baso} {et~al.}(2019{\natexlab{a}}){D{\'\i}az Baso},
      {Mart{\'\i}nez Gonz{\'a}lez}, \& {Asensio Ramos}}]{diazbaso19b}
    {D{\'\i}az Baso}, C.~J., {Mart{\'\i}nez Gonz{\'a}lez}, M.~J., \& {Asensio
      Ramos}, A. 2019{\natexlab{a}}, arXiv e-prints, arXiv:1904.09593
    
    \bibitem[{{D{\'\i}az Baso} {et~al.}(2019{\natexlab{b}}){D{\'\i}az Baso},
      {Mart{\'\i}nez Gonz{\'a}lez}, \& {Asensio Ramos}}]{diazbaso19c}
    {D{\'\i}az Baso}, C.~J., {Mart{\'\i}nez Gonz{\'a}lez}, M.~J., \& {Asensio
      Ramos}, A. 2019{\natexlab{b}}, arXiv e-prints, arXiv:1904.10688
    
    \bibitem[{{D{\'\i}az Baso} {et~al.}(2019{\natexlab{c}}){D{\'\i}az Baso},
      {Mart{\'\i}nez Gonz{\'a}lez}, {Asensio Ramos}, \& {de la Cruz
      Rodr{\'\i}guez}}]{diazbaso19a}
    {D{\'\i}az Baso}, C.~J., {Mart{\'\i}nez Gonz{\'a}lez}, M.~J., {Asensio Ramos},
      A., \& {de la Cruz Rodr{\'\i}guez}, J. 2019{\natexlab{c}}, \aap, 623, A178
    
    \bibitem[{{Du} {et~al.}(2018){Du}, {Lee}, {Li}, {Wang}, \& {Zhai}}]{Du2018}
    {Du}, S.~S., {Lee}, J.~D., {Li}, H., {Wang}, L., \& {Zhai}, X. 2018, arXiv
      e-prints, arXiv:1811.03804
    
    \bibitem[{{Felipe} {et~al.}(2016){Felipe}, {Collados}, {Khomenko}, {Kuckein},
      {Asensio Ramos}, {Balthasar}, {Berkefeld}, {Denker}, {Feller}, {Franz},
      {Hofmann}, {Joshi}, {Kiess}, {Lagg}, {Nicklas}, {Orozco Su{\'a}rez}, {Pastor
      Yabar}, {Rezaei}, {Schlichenmaier}, {Schmidt}, {Schmidt}, {Sigwarth},
      {Sobotka}, {Solanki}, {Soltau}, {Staude}, {Strassmeier}, {Volkmer}, {von der
      L{\"u}he}, \& {Waldmann}}]{Felipe2016}
    {Felipe}, T., {Collados}, M., {Khomenko}, E., {et~al.} 2016, \aap, 596, A59
    
    \bibitem[{{Frutiger} {et~al.}(2000){Frutiger}, {Solanki}, {Fligge}, \&
      {Bruls}}]{Frutiger2000}
    {Frutiger}, C., {Solanki}, S.~K., {Fligge}, M., \& {Bruls}, J.~H.~M.~J. 2000,
      \aap, 358, 1109
    
    \bibitem[{{Gingerich} {et~al.}(1971){Gingerich}, {Noyes}, {Kalkofen}, \&
      {Cuny}}]{HSRA1971}
    {Gingerich}, O., {Noyes}, R.~W., {Kalkofen}, W., \& {Cuny}, Y. 1971, \solphys,
      18, 347–365
    
    \bibitem[{Goodfellow {et~al.}(2016)Goodfellow, Bengio, \&
      Courville}]{Goodfellow2016}
    Goodfellow, I., Bengio, Y., \& Courville, A. 2016, Deep Learning (MIT Press),
      \url{http://www.deeplearningbook.org}
    
    \bibitem[{{Harker} \& {Mighell}(2012)}]{Harker2012}
    {Harker}, B.~J. \& {Mighell}, K.~J. 2012, \apj, 757, 8
    
    \bibitem[{{He} {et~al.}(2015){He}, {Zhang}, {Ren}, \&
      {Sun}}]{residual_network16}
    {He}, K., {Zhang}, X., {Ren}, S., \& {Sun}, J. 2015, ArXiv e-prints
      [\eprint[arXiv]{1512.03385}]
    
    \bibitem[{{Illarionov} \& {Tlatov}(2018)}]{Illarionov18}
    {Illarionov}, E.~A. \& {Tlatov}, A.~G. 2018, \mnras, 481, 5014
    
    \bibitem[{Ioffe \& Szegedy(2015)}]{batchnormalization15}
    Ioffe, S. \& Szegedy, C. 2015, in Proceedings of the 32Nd International
      Conference on International Conference on Machine Learning - Volume 37,
      ICML'15, 448--456
    
    \bibitem[{{Khomenko} \& {Collados}(2007)}]{Khomenko2007}
    {Khomenko}, E. \& {Collados}, M. 2007, \apj, 659, 1726
    
    \bibitem[{{Kingma} \& {Ba}(2014)}]{adam14}
    {Kingma}, D.~P. \& {Ba}, J. 2014, ArXiv e-prints [\eprint[arXiv]{1412.6980}]
    
    \bibitem[{{Lagg} {et~al.}(2004){Lagg}, {Woch}, {Krupp}, \&
      {Solanki}}]{Lagg2004}
    {Lagg}, A., {Woch}, J., {Krupp}, N., \& {Solanki}, S.~K. 2004, \aap, 414, 1109
    
    \bibitem[{{Landi Degl'Innocenti} \& {Landolfi}(2004)}]{landi_landolfi04}
    {Landi Degl'Innocenti}, E. \& {Landolfi}, M. 2004, Polarization in Spectral
      Lines (Kluwer Academic Publishers)
    
    \bibitem[{LeCun \& Bengio(1998)}]{LeCun1998}
    LeCun, Y. \& Bengio, Y. 1998, in The Handbook of Brain Theory and Neural
      Networks, ed. M.~A. Arbib (Cambridge, MA, USA: MIT Press), 255--258
    
    \bibitem[{{L{\"o}ptien} {et~al.}(2018){L{\"o}ptien}, {Lagg}, {van Noort}, \&
      {Solanki}}]{Loptien18}
    {L{\"o}ptien}, B., {Lagg}, A., {van Noort}, M., \& {Solanki}, S.~K. 2018, \aap,
      619, A42
    
    \bibitem[{{Mart{\'\i}nez Gonz{\'a}lez} {et~al.}(2006){Mart{\'\i}nez
      Gonz{\'a}lez}, {Collados}, \& {Ruiz Cobo}}]{2006A&A...456.1159M}
    {Mart{\'\i}nez Gonz{\'a}lez}, M.~J., {Collados}, M., \& {Ruiz Cobo}, B. 2006,
      \aap, 456, 1159
    
    \bibitem[{Nair \& Hinton(2010)}]{relu10}
    Nair, V. \& Hinton, G.~E. 2010, in Proceedings of the 27th International
      Conference on Machine Learning (ICML-10), June 21-24, 2010, Haifa, Israel,
      807--814
    
    \bibitem[{{November} \& {Simon}(1988)}]{NovemberSimon_1988}
    {November}, L.~J. \& {Simon}, G.~W. 1988, \apj, 333, 427
    
    \bibitem[{{Osborne} {et~al.}(2019){Osborne}, {Armstrong}, \&
      {Fletcher}}]{osborne2019}
    {Osborne}, C. M.~J., {Armstrong}, J.~A., \& {Fletcher}, L. 2019, \apj, 873, 128
    
    \bibitem[{{Puschmann} {et~al.}(2010){Puschmann}, {Ruiz Cobo}, \& {Mart{\'\i}nez
      Pillet}}]{Puschmann10}
    {Puschmann}, K.~G., {Ruiz Cobo}, B., \& {Mart{\'\i}nez Pillet}, V. 2010, \apj,
      720, 1417
    
    \bibitem[{{Quintero Noda} {et~al.}(2015){Quintero Noda}, {Asensio Ramos},
      {Orozco Su{\'a}rez}, \& {Ruiz Cobo}}]{Quintero2015}
    {Quintero Noda}, C., {Asensio Ramos}, A., {Orozco Su{\'a}rez}, D., \& {Ruiz
      Cobo}, B. 2015, \aap, 579, A3
    
    \bibitem[{{Rempel}(2012)}]{Rempel12}
    {Rempel}, M. 2012, \apj, 750, 62
    
    \bibitem[{{Riethm{\"u}ller} {et~al.}(2017){Riethm{\"u}ller}, {Solanki},
      {Barthol}, {Gandorfer}, {Gizon}, {Hirzberger}, {van Noort}, {Blanco
      Rodr{\'\i}guez}, {Del Toro Iniesta}, {Orozco Su{\'a}rez}, {Schmidt},
      {Mart{\'\i}nez Pillet}, \& {Kn{\"o}lker}}]{Riethmuller2017}
    {Riethm{\"u}ller}, T.~L., {Solanki}, S.~K., {Barthol}, P., {et~al.} 2017, The
      Astrophysical Journal Supplement Series, 229, 16
    
    \bibitem[{{Rimmele} {et~al.}(2012){Rimmele}, {Keil}, {McMullin}, {Kn{\"o}lker},
      {Kuhn}, {Goode}, {Rosner}, {Casini}, {Lin}, {Tritschler}, {W{\"o}ger}, \&
      {ATST Team}}]{DKIST2012}
    {Rimmele}, T.~R., {Keil}, S., {McMullin}, J., {et~al.} 2012, in Astronomical
      Society of the Pacific Conference Series, Vol. 463, Second ATST-EAST Meeting:
      Magnetic Fields from the Photosphere to the Corona., ed. T.~R. {Rimmele},
      A.~{Tritschler}, F.~{W{\"o}ger}, M.~{Collados Vera}, H.~{Socas-Navarro},
      R.~{Schlichenmaier}, M.~{Carlsson}, T.~{Berger}, A.~{Cadavid}, P.~R.
      {Gilbert}, P.~R. {Goode}, \& M.~{Kn{\"o}lker}, 377
    
    \bibitem[{{Ruiz Cobo} \& {Asensio Ramos}(2013)}]{ruizcobo_asensioramos13}
    {Ruiz Cobo}, B. \& {Asensio Ramos}, A. 2013, \aap, 549, L4
    
    \bibitem[{{Ruiz Cobo} \& {del Toro Iniesta}(1992)}]{sir92}
    {Ruiz Cobo}, B. \& {del Toro Iniesta}, J.~C. 1992, ApJ, 398, 375
    
    \bibitem[{{Socas-Navarro}(2003)}]{socas_nn_03}
    {Socas-Navarro}, H. 2003, Neural Networks, 16, 355
    
    \bibitem[{{Socas-Navarro}(2005)}]{socas_nn_05}
    {Socas-Navarro}, H. 2005, \apj, 621, 545
    
    \bibitem[{{Socas-Navarro} {et~al.}(2015){Socas-Navarro}, {de la Cruz
      Rodr{\'\i}guez}, {Asensio Ramos}, {Trujillo Bueno}, \& {Ruiz
      Cobo}}]{socas_navarro_2015}
    {Socas-Navarro}, H., {de la Cruz Rodr{\'\i}guez}, J., {Asensio Ramos}, A.,
      {Trujillo Bueno}, J., \& {Ruiz Cobo}, B. 2015, \aap, 577, A7
    
    \bibitem[{{Tiwari} {et~al.}(2013){Tiwari}, {van Noort}, {Lagg}, \&
      {Solanki}}]{Tiwari2013}
    {Tiwari}, S.~K., {van Noort}, M., {Lagg}, A., \& {Solanki}, S.~K. 2013, \aap,
      557, A25
    
    \bibitem[{{Uitenbroek} \& {Criscuoli}(2011)}]{Uitenbroek11}
    {Uitenbroek}, H. \& {Criscuoli}, S. 2011, \apj, 736, 69
    
    \bibitem[{{van Noort}(2012)}]{vannoort12}
    {van Noort}, M. 2012, A\&A, 548, A5
    
    \bibitem[{{V{\"o}gler} {et~al.}(2005){V{\"o}gler}, {Shelyag}, {Sch{\"u}ssler},
      {Cattaneo}, {Emonet}, \& {Linde}}]{vogler05}
    {V{\"o}gler}, A., {Shelyag}, S., {Sch{\"u}ssler}, M., {et~al.} 2005, \aap, 429,
      335
    
    \end{thebibliography}
\end{document}